# Title

# Epistatic models predict mutable sites in SARS-CoV-2 proteins and epitopes

Juan Rodriguez-Rivas*[1], Giancarlo Croce*[2,3], Maureen Muscat[1], Martin Weigt[1]

[1] Sorbonne Université, CNRS, Institut de Biologie Paris Seine, Computational and Quantitative Biology – LCQB, Paris, France.

[2] Department of Oncology, Ludwig Institute for Cancer Research Lausanne, University of Lausanne, Switzerland

[3] Swiss Institute of Bioinformatics (SIB), Lausanne, Switzerland

* contributed equally to this work

Correspondence: martin.weigt@sorbonne-universite.fr

## Abstract

The emergence of new variants of SARS-CoV-2 is a major concern given their potential impact on the transmissibility and pathogenicity of the virus as well as the efficacy of therapeutic interventions. Here, we predict the mutability of all positions in SARS-CoV-2 protein domains to forecast the appearance of unseen variants. Using sequence data from other coronaviruses, pre-existing to SARS-CoV-2, we build statistical models that do not only capture amino-acid conservation but more complex patterns resulting from epistasis. We show that these models are notably superior to conservation profiles in estimating the already observable SARS-CoV-2 variability. In the receptor binding domain of the spike protein, we observe that the predicted mutability correlates well with experimental measures of protein stability and that both are reliable mutability predictors (ROC AUC ~0.8). Most interestingly, we observe an increasing agreement between our model and the observed variability as more data become available over time, proving the anticipatory capacity of our model. When combined with data concerning the immune response, our approach identifies positions where current variants of concern are highly overrepresented. These results could assist studies on viral evolution, future viral outbreaks and, in particular, guide the exploration and anticipation of potentially harmful future SARS-CoV-2 variants.

## Significance statement

During the COVID pandemic, new SARS-CoV-2 variants emerge and spread, some being of major concern due to their increased infectivity or their capacity to reduce vaccine efficiency. Anticipating new mutations, which might give rise to new variants, would be of great interest. Here we construct sequence models predicting how mutable SARS-CoV-2 positions are, using a single SARS-CoV-2 sequence and databases of other coronaviruses. Predictions are tested against available mutagenesis data and the observed variability of SARS-CoV-2 proteins. Interestingly, our predictions agree increasingly with observations, as more SARS-CoV-2 sequences become available. Combining predictions with immunological data, we find a clear overrepresentation of mutations in current variants of concern. The approach may become relevant for potential outbreaks of future viral diseases.

# Introduction

The emergence of variants of the severe acute respiratory syndrome coronavirus 2 (SARS-CoV-2) is a major global health concern. Mutations observed in circulating Variants of Interest (VOI) or Variants of Concern (VOC) have been associated with increased transmissibility (1, 2), reduced efficacy of antibody treatments (3, 4), and lower antibody neutralization (5). It is currently under investigation in how far circulating or future mutants can escape the human immune response induced by vaccination or previous infection (6).

Since the beginning of the COVID-19 pandemic, genomic surveillance of SARS-CoV-2 strains has played a pivotal role in tracking new mutations as they appear and expand. Viral sequences sampled from infected individuals from various parts of the world have been continuously deposited in the GISAID database (7), and - as of May 2021 - more than 1,500,000 genomes are available. Genome-wide analysis of circulating strains (8, 9) has shown that mutated positions are heterogeneously distributed across SARS-CoV-2 proteins: while the vast majority of positions has remained largely invariant to date, a restricted set is accumulating diversity. According to Nextstrain (10) global analysis (May 2021, 3883 genomes), no mutational event has occurred for 58% of the entire proteome, while only 14% has experienced more than 2 events. In particular, the protein cores tend to be less variable as mutations in the core usually have a deleterious effect on the stability of the protein (11, 12). In contrast, the exposed regions of the spike protein have accumulated a large number of mutations resulting in variants with increased affinity with the human ACE2 receptor (13, 14), transmissibility (1, 2), and reduced antibody neutralization (5). Each residue of the SARS-CoV-2 proteome is subjected to different selective pressures, which affect the evolution of the virus, thus constraining the variability of SARS-CoV-2 sequences. This suggests that statistical patterns in sequences could be used to distinguish mutable from constrained positions.

In recent years, data-driven models trained on sequence data of patients affected by the human immunodeficiency virus (HIV) have been used in this spirit. They identify regions subject to strong selective constraints and therefore less likely to variate (15, 16), guiding the immunogen design of therapeutic strategies being effective against current and future HIV strains (17, 18). Such approaches are trained on large amounts of HIV sequence data, resulting from decades of study and high rates of intra-patient evolution (19). One of the most important lessons of these studies is the importance of epistasis, *i.e.* the dependence of mutational effects on other pre-existing mutations: epistatic models outperform significantly simpler non-epistatic modeling approaches based on independent conservation patterns of individual residue positions.

The techniques used for studying HIV sequences are not directly applicable to SARS-CoV-2, for which less than two years of data are available, and intra-patient evolution is more limited due to the typically short duration of SARS-CoV-2 infections. We therefore use a strategy, which requires only a single SARS-CoV-2 genome to be known: this genome serves as a reference to build alignments of homologous but diverged sequences, which in our case belong mostly to other coronaviruses. These sequences are used to train statistical sequence models, which in turn can be used to predict the mutability of each position in the proteins of the reference SARS-CoV-2. The advantage of this approach is that predictions rely exclusively on data available very early in the outbreak, and predictions can be tested while more data accumulate.

Current approaches predict the mutability along the SARS-CoV-2 proteome using conservation profiles built from multiple sequence alignments (MSA) of SARS-CoV-2 and other related coronaviruses (20, 21). The resulting models (hereinafter independent models or IND) have few parameters and can be trained using limited sets of data. Unfortunately, they also have only limited predictive power as they assume that positions within a protein evolve independently from each other, disregarding that residues can affect each other's evolution via epistatic interactions.

In this work, we construct unsupervised probabilistic models to predict SARS-CoV-2 mutable and constrained positions. We base our approach on the Direct Coupling Analysis (DCA) (22) that overcomes the before-mentioned limitations by explicitly including pairwise epistatic terms into our modeling. The DCA models are trained using families of homologous sequences, broadly collected from all known coronavirus genomes, allowing us to model the general selective pressures acting on the family of coronaviruses. While the use of other coronaviruses enlarges substantially the datasets making data-driven modeling more robust, we may, however, partially lose information about host-specific constraints like the interaction with host-cell receptors (*e.g.* ACE2 for SARS-CoV-2) or with the host's immune system.

While models are learned from diverged homologs, the prediction of mutable sites requires a SARS-CoV-2 genome (in our case the Wuhan-Hu-1 strain) to be used as reference: our models assign a mutability score to each position in each SARS-CoV-2 protein. This score reflects the constraints acting on a position when mutating away from the reference strain. Other SARS-CoV-2 genomes are only required to test our predictions: we assess the predictive power of our approach and of IND models by validating the predictions with the mutations actually observed in SARS-CoV-2 proteomes deposited in GISAID (gisaid.org, (7)). We carry out a detailed study of the receptor binding domain (RBD) as it plays a pivotal role in viral attachment, fusion and entry and is the primary target for antibody therapies and vaccine development (23, 24). For this specific domain, additional deep mutational scanning (DMS) data are available measuring how amino acid mutations of RBD affect protein expression (a proxy of protein stability) and binding to the human ACE2 receptor (25), allowing us to investigate more deeply their relationship with the DCA mutability score, and with the observed variability across SARS-CoV-2 variants.

Most observed mutations are neutral and do not affect the virus phenotype (26), however mutations occurring in SARS-CoV-2 immunogenic regions, *i.e.* targets of human B and T cells, may allow the virus to evade the immune response induced by vaccination or previous infection. By combining our DCA-mutability scores with data from the Immune Epitope Database (IEDB, (27)), we identify a restricted set of positions in the RBD that are expected to be both mutable and highly immunogenic. Interestingly, we observe that most circulating SARS-CoV-2 VOC or VOI have mutations in a subset of those positions. This combined approach also suggests novel positions that are more likely to mutate in the future and whose mutations could induce a reduction in immune response. In this sense, our predictions may help the rational design of new immunogenic or therapeutic strategies, such as monoclonal antibodies or vaccines, to become more efficient against potential future SARS-CoV-2 strains by targeting less mutable positions.

Data are highly dynamic during the ongoing pandemic. A new variant, Omicron (B.1.1.529), has recently emerged and was rapidly declared VOC during the final revision of this paper. Due to the great interest in characterizing this variant, we have included a new analysis of its RBD mutations. As compared to all pre-existing variants, Omicron increases even the number of mutable and immunogenic positions.

Also, beyond the case of SARS-CoV-2, our study can be seen as a proof-of-concept study. Since we need only a single reference genome to search for distant homologs and make predictions, the approach can be applied very early in any potential viral outbreak in the future. Such predictions may be particularly valuable in situations, where observational data on newly emerging pathogens are missing.

## Results

According to the Pfam protein-domain family database (28), the SARS-CoV-2 proteome (isolate Wuhan-Hu-1) contains 39 protein domains (see Table S1) covering 81% (7860 out of 9748 residues) of the entire proteome. For each of these domains, we predict the mutability using both the epistatic DCA and the independent IND models following the general scheme illustrated in Fig. 1 and detailed in *Materials and Methods*:

- For each protein (domain), we extract MSA of homologous sequences from public sequence databases. These sequences, which belong almost exclusively to other coronaviridae, diverged during up to ~$10^3$-$10^8$ years from their common ancestors with SARS-CoV-2 (29). They are used to train IND and DCA models. These models are applied to the protein sequences of the SARS-CoV-2 reference strain Wuhan-Hu-1 to predict the mutability of each site. Note that only a single SARS-CoV-2 sequence is needed in this step.
- We validate the models using deep-mutational scanning data measuring protein expression, which are currently available only for the RBD of the SARS-CoV-2 spike protein. To this aim, we compare experimentally measured mutational effects with model-based predictions.
- We use SARS-CoV-2 sequences extracted from GISAID to estimate the empirical variability among circulating SARS-CoV-2 strains and to test our predictions. Note that these data are independent from the data used in the first step.

In both approaches, IND and DCA, we thus use the MSA of distant homologs to learn a family-specific sequence landscape $E(a_1, \ldots, a_L)$, or "statistical energy", which provides low values to good (functional), and high values to bad (non-functional) sequences. In this context, $(a_1, \ldots, a_L)$ stands for an aligned sequence, *i.e.* the entries may be any of the 20 natural amino acids or an alignment gap. Any variant containing one or more mutations with respect to the reference sequence in Wuhan-Hu-1, can now be characterized by the statistical-energy change $\Delta E = E(reference) - E(variant)$ assigning *positive values* to variants predicted to be *beneficial*, and *negative values* to variants predicted to be *deleterious*. To obtain a position-specific (but not amino-acid specific) mutability score, we average $\Delta E$ over all amino-acid changes in this position reachable by a single nucleotide mutation, see *Materials and Methods* for the precise definition of $E(a_1, \ldots, a_L)$ and the derived mutability scores.

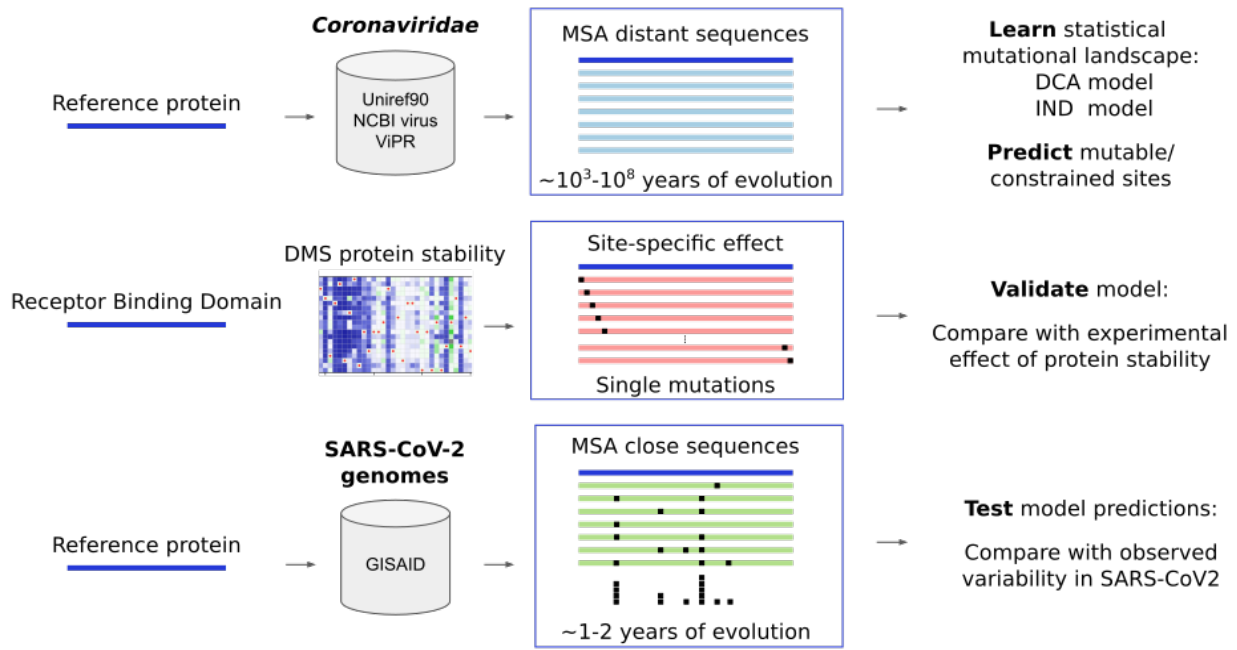

*Fig. 1. Scheme of the protocol and data used in the study. The DCA (epistatic) and IND (independent) models are trained with multiple sequence alignments of diverse sequences coming from large sequence databases. For the receptor binding domain RBD, we add results of deep mutational scanning experiments (DMS) for protein expression (a proxy for stability) (25) which are used as a first independent validation of the models. Model-based predictions are tested against the observed variability, which is derived from SARS-CoV-2 genomes available in GISAID. The estimate on the years of evolution in MSAs of distant sequences provided here (29) is indicative; it can vary strongly between distinct protein domains of SARS-CoV-2.*

For testing these mutability scores, we use the same 39 protein domains for extracting a second MSA with variants of SARS-CoV-2 from the GISAID database. To minimize frequency biases due to the extremely heterogeneous sequencing efforts in different countries, we decided to remove redundant amino-acid sequences and keep each distinct sequence only once. The position-specific observed variability is now defined as the number of distinct sequences in the resulting MSA having a mutation in the position under consideration, when compared to the Wuhan-Hu-1 reference amino-acid sequences (see *Materials and Methods* for more details).

In the case of the RBD, these data (predicted mutability and observed variability) are complemented with the experimental measures for protein expression, used as a proxy for protein stability by Starr *et al. (25)*. This type of data currently exists only for the RBD; we therefore decided to use the RBD for extensive validation of our predictions, and to report predictions for the other 38 domains only at the end of the *Results* section.

**Data-driven sequence landscapes predict mutability and mutational effects in the receptor binding domain**

To assess our prediction framework, we proceed with the following steps. First, we show that the mutational effect predictions of single-site amino-acid mutations are correlated with the protein expression changes measured in the before-mentioned deep mutational scan. Second, we show that the RBD variants available in the GISAID database are significantly more neutral than a randomized sequence library of the same sequence divergence from the Wuhan-Hu-1 reference, the latter showing an accumulation of deleterious mutations. These two observations allow us to use predicted close-to-neutrality of a position as an indicator of its high mutability, while positions predicted to have mostly deleterious mutations are expected to be of low mutability. We use extensive comparisons with the sequence variability derived from GISAID to test this hypothesis.

As a first test, we compare the agreement of the computationally predicted mutational effects with the experimental protein expression. Taking into consideration the region of the RBD domain defined by the Pfam profile *bCoV_S1_RBD* (PF09408, aligned length $L = 178$), we compare our predictions with the experimentally measured single-site mutations. We focus on the *position-specific* mutability, obtained by averaging both predictions and experiments over all accessible amino-acid changes in a position (*Materials and Methods*). The DCA model is well correlated with experimental expression (Fig. 2A, Spearman's ρ=0.54), clearly superior to the IND model (Fig. S1A, ρ=0.32). We also check that individual position and amino-acid specific predictions follow a similar trend (Fig. S1B, ρ=0.49 DCA; Fig S1C, ρ=0.29 IND). In brief, we observe that the protein expression and the epistatic model are well correlated, notably better than for the independent model. We note that the model predicts some mutations to be deleterious, which are neutral in the expression experiments. Two reasons are possible: (a) in limited datasets of functional sequences, undersampled neutral variants may appear deleterious; and (b) mutations without effect on expression may be deleterious for other phenotypes contributing to protein fitness. This observation agrees with what has been observed across other protein families (30–32), where phenotypes better describing fitness also correlate better to sequence-based predictions.

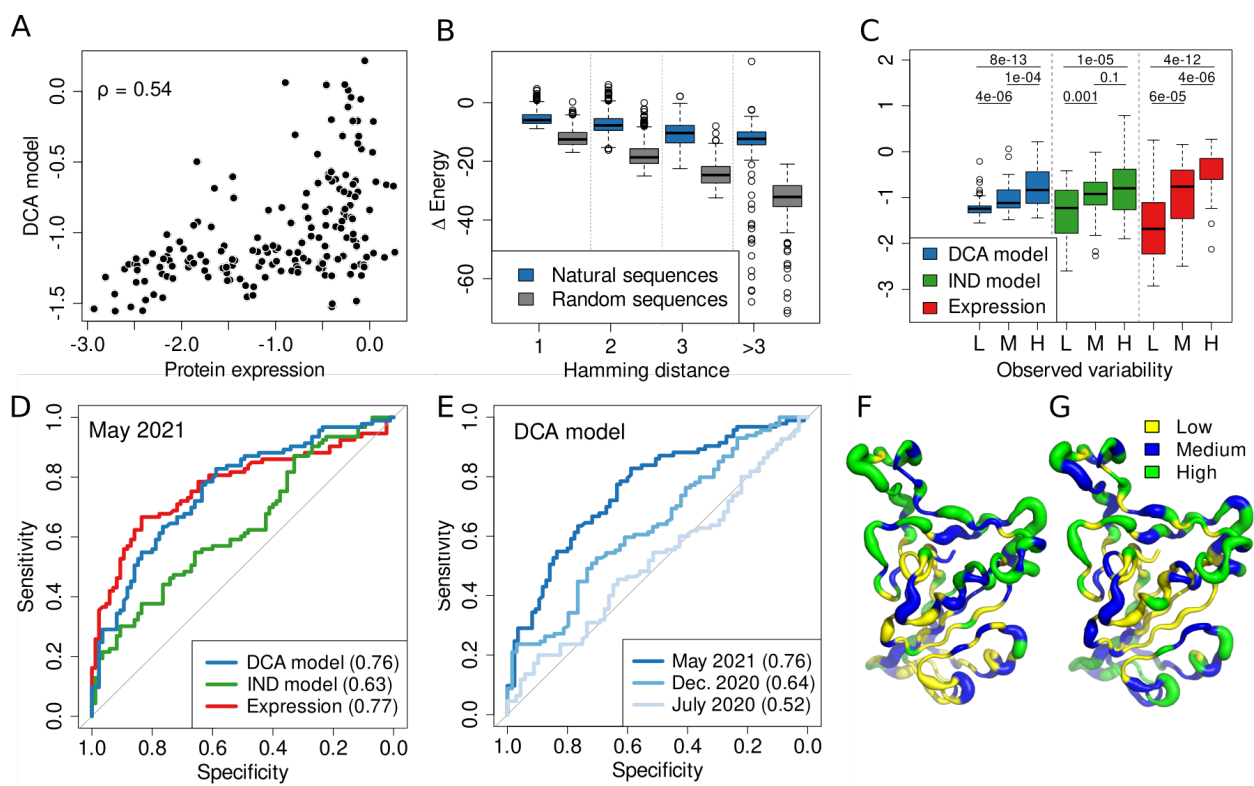

*Fig 2. A)* DCA-predicted mutational scores for the 178 positions of the RBD as a function of the experimental protein expression. *B)* Energy of GISAID sequences and random sequences compared to the reference RBD sequence (from isolate

*Wuhan-Hu-1) indicating how well they fit the DCA model as a function of their Hamming distance. **C)** Distributions of scores from the 3 predictors for positions with low (L, cutoff: <9, n=61), medium (M, [9,16], n=57) and high (H, >16, n=60) observed variability in GISAID. The p-values are obtained with the Wilcoxon signed-rank test. L, M, H in the x-axis correspond to low, medium and high observed variability respectively. **D)** ROC curves for the DCA model for positions with low (cutoff: ≤12, n=93) vs. high (>12, n=85) observed variability for the 3 predictors. **E)** ROC curves for positions with low versus high observed variability, where the observed variability is quantified with the SARS-CoV-2 genomes available at July 2020, December 2020 and May 2021 (Fig. S2A), i.e. with increasing accuracy. **F+G)** RBD 3D structure (PDB code: 6M0J (33)) colored according to 3 levels of mutational scores from the DCA model (panel F) and the protein expression (panel G). Lower mutational scores are shown in yellow, medium in blue and higher in green. The width (wider for higher variability) in both panels corresponds to 3 levels of the observed variability (same cutoffs as in panel C). In all ROC curves, the area under the curve (AUC) of the ROC is shown in the legend. Cutoffs to define positions with low and high variability in the ROC analyses were chosen to split into balanced subsets with the most similar number of observations possible in each subset (Fig S2A). i.e. the median is used as the cutoff. ROC curves for the other combinations of predictors and dates are shown in Fig S3.*

Our DCA and IND models are built from MSAs of diverged species, where we explicitly remove sequences similar to the Wuhan-Hu-1 reference (*Materials and Methods*) to avoid overlaps with the GISAID sequences used to estimate the local SARS-CoV-2 observed variability. We compare how fit, according to the DCA model, the natural sequences are compared to sequences having the same number of random mutations. This can be achieved by comparing the statistical-energy differences $\Delta E = E(reference) - E(variant)$ between the two sets of sequences. Fig. 2B shows that the natural SARS-CoV-2 variants are significantly better according to the model than randomly mutated sequences, *i.e.* the naturally occurring mutations are, according to the model, significantly more neutral than the predominantly deleterious random mutations, as is to be expected by evolution under purifying selection. This finding indicates the capacity of DCA trained on diverged homologs to capture local constraints acting on the evolution of SARS-CoV-2 proteins.

The combination of these two observations is key for our work: the epistatic model is able to capture mutational effects, and the SARS-CoV-2 variants, which emerged over the last months, are significantly more neutral than random mutations. Can we use this to predict possible new variants of SARS-CoV-2 by identifying positions with favorable mutability scores? To test this idea, we compare the currently observable SARS-CoV-2 variability with the one predicted using the model-based mutability score, and with the mutations expectable by the experimental protein expression. As mentioned before and illustrated in Fig. 1, we operationally assess the observable variability by the number of distinct GISAID sequences having a variant amino acid (compared to the reference Wuhan-Hu-1) in the specific position under study. We observe that the DCA model and expression are similarly correlated with variability (Spearman's ρ=0.61 and 0.6, respectively), while the correlation is weaker for the IND model (ρ=0.34). This trend can also be observed in Fig. 2C by looking at the distribution of mutational scores after grouping positions by their observed variability, they grow accordingly with the variability (note that the scores are not comparable between methods, but p-values indicate a higher significance for DCA and expression than for IND).

We analyze in detail the performance of these different measures as a predictor of SARS-CoV-2 mutability through Receiver Operating Characteristic curves (ROC curves) and the resulting areas under curve (AUC), which range from 0.5 for random to 1.0 for perfect predictions. We perform the ROC analysis using a variability cutoff of 12, because it splits the set of positions into two balanced subsets of positions with low (≤12, n=93) or high (>12, n=85) variability. The DCA model and protein expression (AUC 0.76 and 0.77) show a remarkable performance in distinguishing positions with low or high variability, clearly outperforming the IND model (AUC 0.63). This result is not dependent on the particular cutoff chosen, as a similar trend is observed for a large range of variability cutoffs (Fig. S2B). The protein expression and the DCA model perform similarly (see Fig. S2B, averaged ROC AUC of 0.83 and 0.81, respectively), followed by the IND model (0.66). The experiment-based predictor performs comparatively better at high variability while the sequence-based ones are better at low variability (Fig. S2B), which is probably related to the fact that highly conserved positions are usually very relevant to the function of the protein (34). In conclusion, we observe that the different measures have a substantial predictive power of the mutability, although the IND model is worse compared to the others. The performance of the DCA models is surprisingly competitive, with a similar performance as the experimental measurements, with an advantage for lower mutabilities and a slight disadvantage for higher mutabilities, cf. Fig. 2B.

The performance measured in the previous analysis is not only dependent on the intrinsic predictive power of each method but how well the ground truth is defined, in this case the variability estimated from GISAID data. Although only a fraction of all possible variants emerged in the short time since SARS-CoV-2 appeared, the observed variability has greatly evolved over time due to the great effort of sequencing SARS-CoV-2 genomes (Fig. S2A). As an example, the number of RBD positions without observed variants has shrunk from 58 out of 178 in July 2020, to only 3 in May 2021. Interestingly, we observe a great increase in predictive performance when evaluated against more recent and richer variant libraries considered in the estimation of variability. As shown in Fig. 2E, the performance has increased from an AUC of only 0.52 considering sequences collected until July 2020 up to 0.76 with sequences until May 2021 (at each timepoint, the median variability is used to partition data into low vs. high variability, cf. Fig. S2B and *Materials and Methods* for details). This improvement is not only evident from the ROC analysis, but also by looking at the correlation between the DCA model and the variability with Spearman correlations of $\rho=0.13$, 0.35 and 0.61 in July 2020, December 2020 and May 2021, respectively, cf. Fig. S4. This result indicates that the remarkable increase of SARS-CoV-2 genomes has led to a much better estimation of the variability. More importantly, it shows that our computational model is able to anticipate future variability; it suggests that the performance could be even higher with more data leading to an even better estimation of the variability. We have complemented this analysis also for mutability predictions done using the IND score or protein expression (Fig. S3), all show the tendency to improve with the increase of SARS-CoV-2 data available in the GISAID database. Only the IND-score based analysis shows no clear trend between December 2020 and May 2021.

Recently deep learning has been used to improve mutational predictions in the case of large training MSA, but the accuracy for viral proteins remains rather limited (35). To explore this issue, we have used DeepSequence (35). Its predictions correlate well with the DCA model predictions ($\rho=0.6$, Fig. S5C), but the correlations with protein expression (0.31, Fig. S5A) and observed variability (0.35, Fig. S5D) are smaller than those of the DCA model ($\rho=0.54$ and 0.61), congruent with the prior observations for other viral proteins.

To conclude the comparison of computational predictions, observed variability and experimental DMS data for the RBD, we explore how these are distributed within the three-dimensional RBD structure (Figs. 2F and 2G). Apart from an overall agreement between the three quantities, there is a clear trend for lower values in the core of the RBD and higher in the exposed parts of the structure, probably related to the greater impact of mutations on the stability in the core, and the selective pressure for immune escape in the surface.

As a summary of this Section, we conclude that the epistatic DCA prediction for the position-specific mutability of RBD positions in SARS-CoV-2 is highly informative about the observable variability across the increasing number of sequenced SARS-CoV-2 variants. The increased accuracy when compared to the most recent versions of GISAID proves the anticipatory power of our approach: the positions that are predicted as mutable by our approach are more likely to be associated with future SARS-CoV-2 variants.

**Combining mutability predictions with immune response frequencies identifies mutations present in several SARS-CoV-2 variants of concern**

Nonsynonymous mutations of SARS-CoV-2 occurring in immunogenic regions can cause the virus to (partially) escape the human B and T immune response induced by vaccination or previous infection. B and T cells target specific regions of the viral proteome, known as B/T cell epitopes. Epitope mutations can reduce the ability of the immune cells to recognize and bind epitopes, and thus the effectiveness of the immune response.

Antibody-escaping mutations are already present in circulating variants (4, 36). On May 18th, 2021 the WHO weekly epidemiological report on Covid-19 (available at https://www.who.int/publications/m/item/weekly-epidemiological-update-on-covid-19---18-may-2021) classified six SARS-CoV-2 variants as *variants of interest* (VOI) and four – posing an increased risk to global public health – as *variants of concern* (VOC). Both VOIs and VOCs are likely to affect transmission, diagnostics, therapeutics, or immune escape (37). Within the RBD domain, only 7 positions of VOIs and VOCs strains are mutated with respect to the Wuhan-Hu-1 reference strain (Fig. 3A).

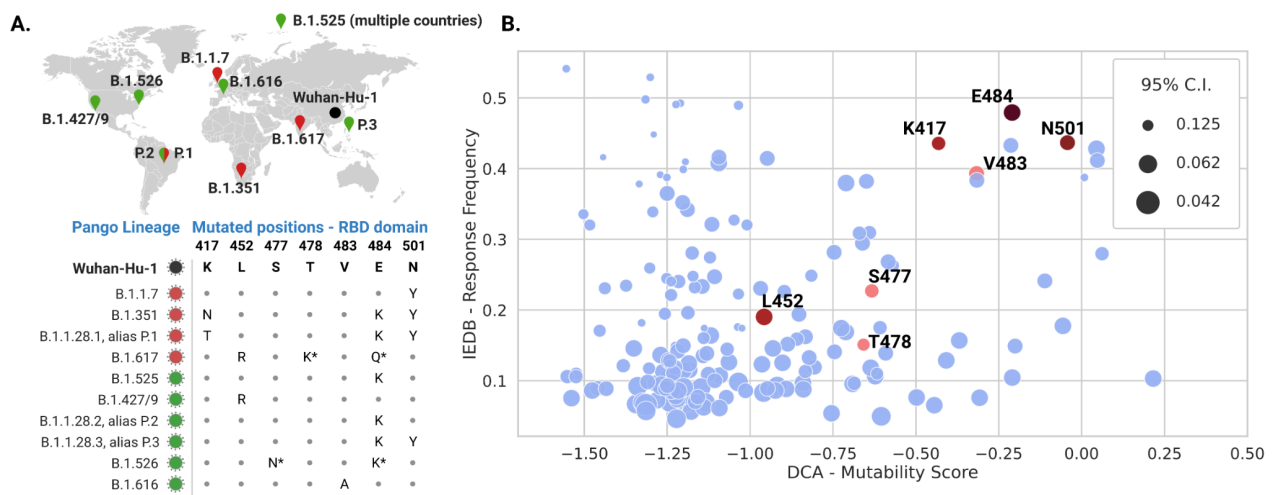

***Fig. 3: A)*** *SARS-CoV-2 strains classified in May 2021 as variants of concern (VOC, red, now also named Alpha (B.1.1.7), Beta (B.1.351), Gamma (P.1) and Delta (B.1.617.2)) and variants of interest (VOI, green). The figure shows the corresponding amino-acid mutations with respect to the Wuhan-Hu-1 reference in the RBD domain and the geographical area where they were first detected. The B.1.617 lineage is divided into three sub-lineages; the E484Q and T478K (with asterisks) mutations are not shared by all sub-lineages. The same is true for E484K and S477N in the B.1.526 lineage.* ***B)*** *The IEDB-Response Frequency and the DCA mutability score for each position of the RBD domain. The upper right corner contains potentially dangerous positions, as they are predicted to be mutable (high DCA mutability score) and are shared by multiple positively responding epitopes (high IEDB-RF). Mutated positions observed in VOCs and VOIs strains are depicted in red, and darker shades correspond to the most frequent mutations. The size of each point is inversely proportional to the IEDB 95%-confidence interval (size ~ 1/(upperbound-lowerbound)), thus larger points correspond to more statistically reliable IEDB-RF.*

The Immune Epitope Database (IEDB, (27)) collects experimentally validated B and T cell epitopes. Most of the SARS-CoV-2 epitopes are localized on the spike protein: a total of 913 epitopes, of which 459 B cell, and 463 T cell epitopes, as of 16 May 2021.

Each position of the spike protein is associated with a site-response frequency (RF), cf. *Materials and Methods*. The RF is calculated as the number of positively responding subjects relative to the total number of those tested, averaged over all epitopes mapped to that position. Nonsynonymous mutations in position with high RF typically modify multiple epitopes with the risk of negatively affecting the human immune response.

In Fig. 3B, we plot the IEDB - RF versus the DCA mutability score for each position of the SARS-CoV-2 RBD domain. Interestingly, only a restricted set of positions has high DCA and RF scores at the same time (the upper right corner of Fig. 3B), and 4 of them are observed in circulating VOCs and VOIs, including the well-known positions N501 and E484. Figs. S6 shows analogous plots of the RF vs. the IND score or the expression data; the enrichment of VOC/VOI mutations becomes less pronounced as compared to the DCA score. These results highlight the potential for our approach to identify positions that are likely to mutate (high DCA score) and whose mutations may cause immune escape (high IEDB - RF). The first 20 predictions, sorted according to the DCA mutability score, are given in Table 1. Note that 9 predictions have a high RF (>0.3, highlighted in bold), and several of them are not yet part of current VOCs and VOIs. Our results suggest them as potentially dangerous positions likely to mutate in future SARS-CoV-2 strains.

As the virus is constantly changing through mutation, other circulating SARS-CoV-2 VOCs/VOIs have emerged during the redaction of the paper. Also, more epitopes have been tested and immunological data are rapidly accumulating, statistically more reliable IEDB RFs are now available. In Dec 2021 we repeated the same analysis using updated IEDB data (download 22 Nov 2021) and the 5 current VOCs. The results reported in the SI in Fig. S7 confirm the enrichment of dangerous mutations in the upper right corner, which is particularly pronounced for the newly emerged Omicron (B.1.1.529) variant. Indeed, of the 14 RBD mutations present in Omicron, 6 (K417, N440, E484, Q493, Q498, N501) are in the first top 20 DCA predictions (Table 1). Remarkably, mutations in positions N440, Q493 and Q498 occur for the first time in Omicron; they are not shared by other VOIs and VOCs.

*Table 1.* The first 20 predictions, sorted according to the DCA mutability score, with the corresponding IEDB - RF, and the VOIs and VOIs in which the position has mutated. In bold, we show positions with IEDB - RF above 0.3

| Position | AA Wuhan-Hu-1 | DCA-Mutability Score | IEDB-Response Frequency (95% CI) | Pango lineage (ref. 38) |
|---|---|---|---|---|
| 519 | H | 0.22 | 0.10 (0.08:0.14) | |
| 403 | R | 0.06 | 0.28 (0.24:0.32) | |
| **490** | **F** | **0.05** | **0.41 (0.38:0.45)** | |
| **493** | **Q** | **0.04** | **0.43 (0.40:0.46)** | |
| **372** | **A** | **0.01** | **0.39 (0.32:0.46)** | |
| **501** | **N** | **-0.04** | **0.44 (0.40:0.47)** | **B.1.1.7; B.1.351; P.1; P.3** |
| 445 | V | -0.06 | 0.18 (0.15:0.21) | |
| 498 | Q | -0.11 | 0.24 (0.21:0.28) | |
| 441 | L | -0.20 | 0.15 (0.12:0.19) | |
| 440 | N | -0.21 | 0.10 (0.08:0.14) | |
| **484** | **E** | **-0.21** | **0.48 (0.45:0.51)** | **B.1.351; P.1; B.1.617; B.1.525; P.2; P.3** |
| **486** | **F** | **-0.21** | **0.43 (0.40:0.47)** | |
| 443 | S | -0.31 | 0.08 (0.05:0.11) | |
| **494** | **S** | **-0.32** | **0.38 (0.35:0.42)** | |
| **483** | **V** | **-0.32** | **0.39 (0.36:0.43)** | **B.1.616** |
| 460 | N | -0.37 | 0.16 (0.13:0.19) | |
| 444 | K | -0.41 | 0.13 (0.10:0.16) | |
| **417** | **K** | **-0.43** | **0.44 (0.40:0.48)** | **B.1.351; P.1** |
| 439 | N | -0.44 | 0.07 (0.04:0.10) | |
| 402 | I | -0.50 | 0.08 (0.05:0.11) | |

We repeated our analysis distinguishing between T and B cell epitopes (Fig. S8). While for B cell epitopes it is still possible to clearly identify a subset of positions with high DCA and RF scores, this is not the case for T cell epitopes. This is expected as B cell-antibodies directly bind the pathogen, while T cell epitope must be presented by the human leukocyte antigens (HLAs) - one of the most polymorphic genes in the human genome - and have a much larger sequence variability. Limited T cell data makes it arduous to obtain a statistically reliable T cell IEDB-Response Frequency, even after restricting the analysis to a subset of T cell epitopes shared by a large fraction of the population (Fig. S9, (39)).

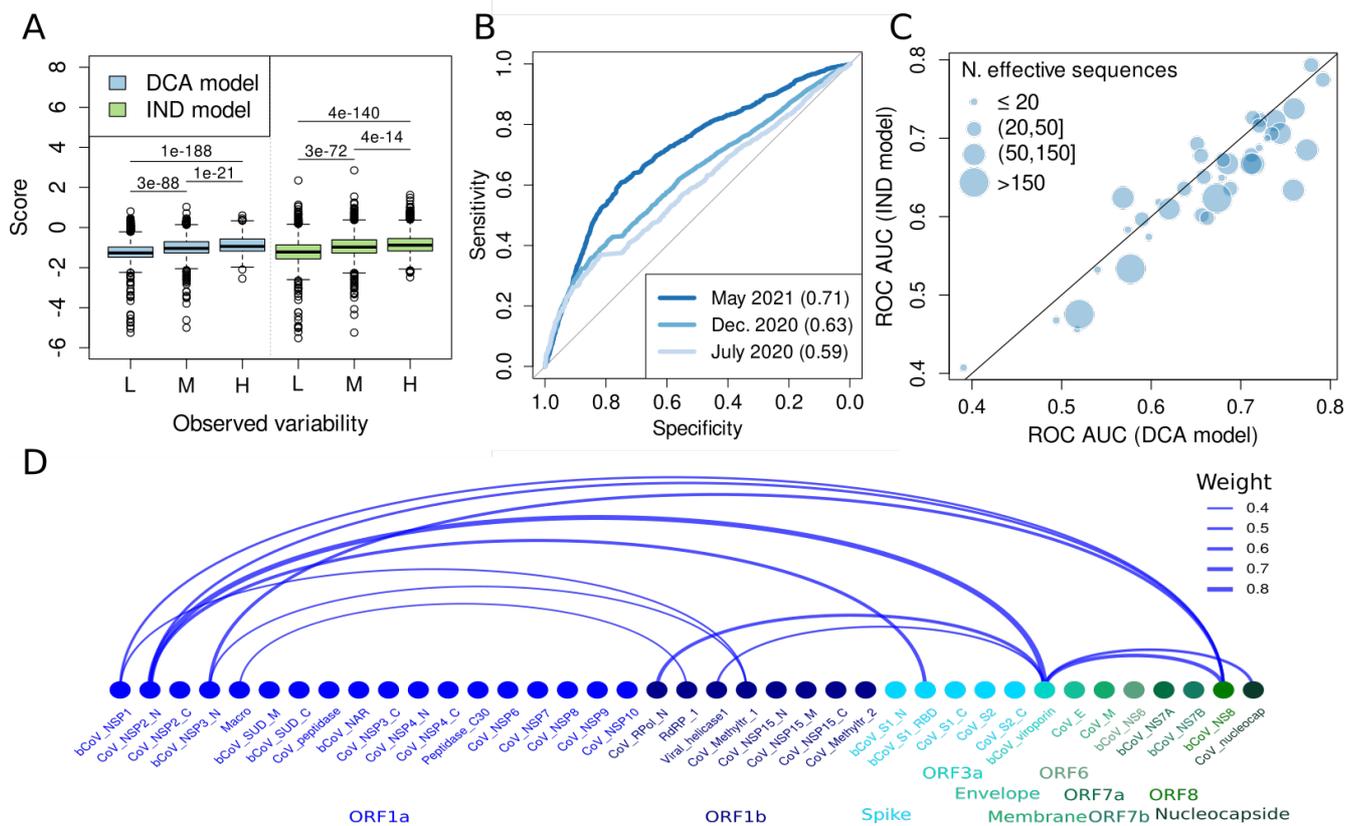

*Fig 4. A)* Distribution of DCA and IND scores as a function of the variability (L, low<7, n=2757 positions; M, medium=[7,15], n=2647; H, high>15, n=2554) for the entire SARS-CoV-2 proteome (p-values from the Wilcoxon signed-rank test). L, M, H in the x-axis correspond to low, medium and high observed variability respectively. *B)* ROC curve for the classification provided by the DCA model for positions with low (≤ 3, n=4873 in December 2020) or high (>3, n=3085 in December 2020) variability, where the variability is estimated from data until May 2021, December 2020 or July 2020. *C)* Comparison of ROC AUC obtained by the DCA and IND models for the 39 domains in the proteome. The variability cutoff for each domain is chosen to give rise to two balanced subsets of positions. *D)* The nodes represent the Pfam domains in the proteome with a link between pairs of domains when they have at least one relatively strong epistatic coupling. The width of the link is proportional to the strength of the signal, or weight, which comes from the strongest coupling among all the inter-domain pairs of positions. Protein domains codified within the same ORF share the same color.

**Mutability predictions are extendable to all SARS-CoV-2 proteins**

Thanks to the wide availability of sequence data as compared to experimental data, a key advantage of our data-driven modeling approach is the possibility to obtain predictions for all the protein domains in the SARS-CoV-2 proteome. We extend the analysis to all 39 protein domains covering 81% of the entire proteome (8037 out of 9748 positions). First, we observed that both the mutational scores from DCA and IND models systematically grow with the observed variability (Fig. 4A), with a more pronounced change for the DCA model reflected by smaller p-values. The performance is influenced by both the quality of the models, which depends on the available sequence data, and the definition of the ground truth, *i.e.* the observed variability. In accordance with previous findings for the RBD, we observe that the performance of the DCA model improves as more data is used to estimate the variability, the ROC AUC goes from 0.59 in July 2020 to 0.71 in

May 2021. The increased performance observed in the RBD can be partially attributed to a better estimation of the variability, as it is one of the most variable regions of SARS-CoV-2 proteome (40). Another important factor is the impact of the available sequence data (Table S1). To take into consideration this factor, we split the 39 protein domains into two sets with at least 50 or fewer than 50 effective sequences (*i.e.* non-redundant at 80% identity, see *Materials and Methods*). As expected, the performance is greater when more sequence data is available to build the models as shown in Fig. S10. A systematic comparison between the models reveals that the DCA model is better than the IND in most cases, especially in those with a higher number of effective sequences (Fig. 4C).

Beyond the RBD domain, other protein domains have an important role in triggering an immune response in humans. We extend the analysis of the previous section (combining immunological data with DCA predictions) to all the protein domains of the SARS-CoV-2 proteome. The results are available on the GitHub page.

Our predictions are based on individual Pfam protein domains. While we argued that epistasis is a crucial ingredient to our models, we currently do not include epistasis between distinct domains. The main reason is that multi-domain studies risk again limiting available sequence data. To get a first impression of the potential role of epistasis between distinct protein domains in SARS-CoV-2 evolution, we have used DCA to detect epistatic couplings between all 741 pairs of the 39 present domains (SI Text). As is shown in Fig. 4D, we find a sparse network of only 12 potentially coupled domain pairs, out of the 601 pairs providing sufficient data for our analysis (SI Text). While the sparsity of this network makes it unlikely that our mutability predictions suffer from our domain-centric modeling approach, our results suggest the existence of inter-domain and inter-protein epistasis in SARS-CoV-2. This conclusion is coherent with the one of (41), which in difference to our analysis is based entirely on an analysis of the SARS-CoV-2 genomes deposited in GISAID. It is worth mentioning that the strongest epistatic coupling found in our analysis is between the domain Cov_NSP2_N and bCov_viroporin (Table S3), which is also highlighted in (41). However, a biological interpretation of these findings is not obvious due to the limited availability of experimental information about potential physical or functional interactions between SARS-CoV-2 proteins.

## Discussion

In this work, we propose to use statistical models to predict the mutability of individual positions in SARS-CoV-2 proteins. The models are based on MSAs coming from various coronaviruses. The inclusion of epistasis into the DCA-based modeling framework allows us to capture local evolutionary constraints specific to the SARS-CoV-2 sequence background. Using several tests for the RBD of the spike protein, for which the most extensive experimental datasets are available, we were able to establish that our computational predictions are able to anticipate position mutated in variants of SARS-CoV-2 from sequence alignments not containing SARS-CoV-2 sequences. This fact is particularly evident in Fig. 2E, which shows that more recent and thus richer releases of the GISAID database of SARS-CoV-2 genomes follow more accurately our model predictions. The inclusion of epistasis into the modeling was found to be essential to improve the quality of the mutability predictions.

The combination of our predictions with available immune response frequencies allows for identifying a relatively small group of 9 positions out of the 178 positions in the RBD, which are highly mutable and have a high potential for immune escape. Interestingly, 4 out of these 9 positions are mutated in the current VOC or VOI. The other 5 positions are predicted to potentially give similar advantages for emerging SARS-CoV-2 variants. In fact, a new variant has been declared VOI in June 2021 (lineage C.37, or Lambda). This variant has two RBD mutations in positions L452 and F490. While the first is shared with other lineages (even if substituted to another amino acid), the second one was not part of any previous VOI or VOC, but it is the third predicted position in terms of mutability and the first with high response frequency (cf. Table 1). Even more recently, in November 2021, the variant Omicron (B.1.1.529) emerged and was declared VOC; it shows new mutation in positions Q493, Q498 and N440, which were not mutated in pre-existing VOIs and VOCs but take ranks 4, 8 and 10 in Table 1, cf. also Fig. S7. Our approach therefore highlights the importance of monitoring these positions, which could also be taken into account when exploring potential therapeutic or vaccine targets.

This can be illustrated by the following example: a monoclonal antibody was recently isolated, which has neutralizing activity against all SARS-CoV-2 VOCs identified to date (42). The antibody targets a region of about 600 $Å^2$ of the spike protein surface centered in residue F486. This residue is predicted to be quite mutable (rank 12 in Table 1, next to E484), and mutations might have immunological relevance as indicated by a high IEDB response frequency – so a mutation in F486 emerging in a new variant might decrease the neutralizing capacities of the antibody. However, this residue is also in contact with the ACE2 human receptor, thus a mutation might also decrease the affinity with the host protein resulting in an evolutionary disadvantage for the virus. While our model, trained on long-term evolutionary data, does not contain specific knowledge about the RDB-ACE2 interaction, it suggests that positions like F486 should be carefully studied with complementary structural and experimental approaches, and considered when designing antibodies effective against novel strains.

Being based on readily available sequence data is one of the advantages of our approach over more labor-intensive experimental approaches like the deep mutational scanning data, such as the effect of mutations on RBD expression and binding to the human ACE2 receptor, allowing us to provide useful predictions of mutability for most of the SARS-CoV-2 proteome. It also has its limitations, most importantly their dependence on the availability of sufficiently large and diverged sequence ensembles. In fact, we observe that a greater number of sequences usually increases the performance of the approach (Fig. 4C and S10). However, it is important to note that the inclusion of more divergent sequences might not always be the best strategy as the model might capture constraints that are not relevant for the specific SARS-CoV-2 context. This trade-off will be explored in future work.

Our approach can be extended in several ways. One is to include how different domains might constrain the variability of other domains. However, according to our analysis in the previous section, inter-domain epistasis seems to play only a minor role, even if more sequence data might be needed to better estimate the influence of inter-domain or inter-protein epistasis. Another is to model constraints due to specific virus-host interaction, which is currently out of our scope, as we do not consider host sequences in the MSAs. Indeed, we observe the correlation of experimental binding to ACE2 and our predictions (Pearson's r = 0.27) can be fully explained through the protein expression (Pearson's r partial correlation controlled by expression = -0.02). In an attempt to explore this issue, we built co-alignments of receptor-binding domains with homologs of ACE2 present in the hosts of other coronaviruses. Since the binding mechanism between RBD and ACE2 homologs is present only in sarbecoviruses, the resulting co-alignment of RBD and ACE2 homologs contains only

7 effective sequences (non-redundant sequences at 80% identity, cf. *Materials and Methods*), a number being insufficient to capture the complex virus-host interactions.

Predicting evolution is an undoubtedly daunting task (43, 44). While there is little, if any, hope to predict specific future evolutionary events, we have shown that data-driven approaches capturing statistical patterns in sequence data can effectively identify more general evolutionary trends, such as which positions are more likely to mutate and represent a concern to current therapeutic interventions. In this sense, our work is a step forward to a more precise characterization of the SARS-CoV-2 evolution fuelled by a huge worldwide effort of research and monitoring of the virus, whose evolution is unfolding in almost real time at an unprecedented level of detail.

While the main application of our work is the insights provided on SARS-CoV-2, our study can also be seen as a proof of concept. In the case of emergence of a new viral pathogen, a single sequenced genome can be used as the reference to first extract families of homologous sequences from public databases, which allow for learning the statistical models needed for mutability predictions. These predictions can therefore be done in very early stages of a possible outbreak, before large amounts of observational or experimental data become available, forecast future variability and thereby help to direct our attention to not yet observed mutations.

## Materials and Methods

### Sequence data

Sequence data in FASTA format were downloaded from the following databases: GISAID ((7), release 16 May 2021), Uniref90 ((45), release December 2020), ViPR ((46), downloaded in September 2020), NCBI viral genomes ((47), downloaded in September 2020) and MERS coronavirus database ((48), downloaded in September 2020). The amino acid sequence of isolate Wuhan-Hu-1 was used as the reference proteome (genbank identifier: MN908947). Protein domains were detected using the HMMER suite ((49), version 3.1b2) and the HMM profiles from Pfam.

A global database including distant species was built by combining Uniref90, ViPR, NCBI viral genomes and MERS coronavirus database, and used to train the DCA and IND models. We built MSAs by running *jackhmmer* with 5 iterations and starting both with the full-length reference protein sequence (except for the ORF1ab) and with the trimmed domain sequence (*SI text* for more details). For each domain, we selected the MSA with more non-redundant sequences between the two resulting MSAs for further analysis, which increases the amount of available sequence data. As quality controls, all sequences including non-standard amino acids were removed as well as repeated sequences or sequences covering less than 80% of the reference; predictions are robust when modifying this threshold, cf. Fig. S11A. To separate training from test data, all sequences closer than 90% sequence identity to the Wuhan-Hu-1 reference were filtered out (*i.e.* all SARS-CoV-2 sequences, including close relatives in non-human hosts). The exclusion of SARS-CoV-2 reference sequences has a negligible influence on the predictions, *e.g.* the spearman's correlation on the RBD of the DCA scores with protein expression with and without the reference sequence is the same ($\rho=0.54$) as well as in the case of the observed variability ($\rho=0.61$).

For the GISAID database, a MSA for each domain sequence was built with only 1 iteration in *jackhmmer* as the GISAID sequences are very similar to the reference sequences. We applied the same quality controls as before but keeping sequences closer than 90% sequence identity and removing sequences corresponding to a non-human host. The July and December 2020 subsets of sequences were collected until the 16th of the corresponding month. The alignments of GISAID sequences were used exclusively for testing our predictions.

The random sequences in Fig. 2B are generated by randomly selecting a position and variant following an uniform distribution. For each GISAID sequence, a random sequence is produced with the same number of mutations to the reference.

## Statistical models

For each protein domain in the reference proteome, we built an independent-site model (IND) and an epistatic model (DCA) using the previously described global MSA containing a diverged set of species, but no SARS-CoV-2 sequences.

*Independent or sequence profile model* – Assuming statistical independence of positions, a simple probabilistic model $P_{IND}(a_1, a_2, \ldots, a_L)$, where $(a_1, a_2, \ldots, a_L)$ represents an aligned sequence of amino acids (with the gaps '-' to accounting for insertions or deletions) of length $L$, for a protein family is defined by

$$P_{IND}(a_1, \ldots, a_L) = \prod_{i=1}^{L} f_i(a_i).$$

The factors $f_i(a)$ equal the empirical frequencies of amino acid $a$ in column $i = 1, \ldots, L$ of the global MSA (with $L$ columns). Therefore, the probability that any sequence of length $L$ belongs to the protein family is factorized into the individual position-specific contributions of each of its amino acids. Similar to (30), the effect of an amino-acid mutation $a_i \to b$ can be computed as

$$\Delta E(i, b) = \log P_{IND}(a_1, \ldots, a_i, \ldots, a_L) - \log P_{IND}(a_1, \ldots, b, \ldots, a_L) = \log f_i(a_i) - \log f_i(b)$$

In contrast to previous work (30), positive values correspond to beneficial mutations while negatives correspond to deleterious mutations. Therefore, this value can be more naturally interpreted as a proxy of the selective pressure acting across coronaviruses.

*DCA or epistatic model* – It is possible to overcome the assumption of independence between positions by introducing two-site coupling terms as done in DCA models:

$$P_{DCA}(a_1, \ldots, a_L) = \frac{1}{Z} \exp\left( \sum_{1 \leq i \leq L} h_i(a_i) + \sum_{1 \leq i < j \leq L} J_{ij}(a_i, a_j) \right)$$

where $Z$ is a normalization constant. The inference of model parameters is a computationally hard task and a number of approximations have been proposed (50–53). In this work, we rely on the widely used asymmetric plmDCA approach (52), which provides one of the best trade-offs between computational cost and performance. Following standard practice (50), we apply a sampling correction by counting the number of sequences with higher than 80% identity and reweighting them (results are robust to the specific value of this parameter, see Fig. S11B). The number of effective sequences refers to the number of sequences that are not redundant at 80% sequence identity. As before, the effect of a single mutation $a_i \to b$ can be computed as the difference between a wildtype sequence and single mutant sequence

$$\Delta E_{DCA}(i,b) = \log P_{DCA}(a_1,\ldots,a_i,\ldots,a_L) - \log P_{DCA}(a_1,\ldots,b,\ldots,a_L)$$

As we focus on the mutability of each position in the SARS-CoV-2 proteome, for each of the models IND and DCA we derive a single mutational score $S_{IND/DCA}$ for each position $i$ as

$$S_{IND/DCA}(i) = \frac{1}{q}\sum_1^q \Delta E_{IND/DCA}(i,b_k)$$

where $\Delta E(i,b_k)$ is the effect of the $k$th single mutations ($a_i \to b_k$) in position $i$. We restrict the set of amino acids to the ones reachable by a single nucleotide missense mutation from the corresponding codon in the Wuhan-Hu-1 reference genome (*i.e.* the alphabet size $q$ depends on the specific codon used in position $i$). To make the quantification more interpretable and comparable between distinct domains, we divide the mutational score by the average score considering all the positions in the domain

$$MS_{IND/DCA}(i) = S_{IND/DCA}(i) / \sum_1^L S_{IND/DCA}(i)$$

This final mutational score is positive for beneficial mutations and negative for deleterious mutations. Values close to 0 can be interpreted as neutral, values in the range (-1,0) as better than average and lower than -1 as worse than average and more deleterious.

### Estimating variability of SARS-CoV-2 sequences from GISAID data

The variability of each position was estimated by counting the number of sequences that have a different amino acid in the corresponding position compared to the reference. Only non-identical sequences were considered to avoid the strong sequencing bias due to the highly diverse number of genomes sequenced in different countries. This corresponds to the standard reweighting procedure used in DCA, but at a 100% similarity threshold adapted to the high sequence similarities between SARS-CoV-2 strains. Results are robust with respect to this procedure, since the empirical variability estimated without any reweighting shows, in the RBD, a Pearson correlation of 0.89 (Spearman correlation of 0.92) to the reweighted estimates.

### Deep mutational scanning data

The deep mutational scanning data measuring protein expression and the binding to ACE2 obtained by Starr *et al. (25)* was collected from https://github.com/jbloomlab/SARS-CoV-2-RBD_DMS trimmed to the RBD alignment (which contains 178 Pfam positions instead of the 201 in the experiment) and merged into our framework.

### IEDB data

B and T cells epitope data were collected from the IEDB webserver by selecting Organism SARS-CoV-2 (ID:2697049, SARS2), and restricting to B and T cells assay. For each protein of the SARS-CoV-2 proteome, a list of experimentally validated epitopes is provided. Following the definition of https://help.iedb.org/hc/en-us/articles/114094147751, it is possible to introduce a Response Frequency (RF) for each position $i$ of the proteome. RF is defined as the number of positively responding subjects relative to the total number of those tested, averaged over all epitopes mapped to that position. Large values thus correspond to positions of high potential for immune escape.

The IEDB website only reports the upper and lower bounds of the 95% confidence interval (CI) for the RF score – and not the RF score itself – to correct for the sample size. In our analysis we compute the mean RF score from the IEDB epitope data and use the 95% confidence upper and lower bounds provided by the IEDB to compute the confidence interval (CI = upperbound-lowerbound) for each position.

**Performance evaluation**

All ROC analyses were performed in R ((54), version 3.6.3) using the package pROC ((55), version 1.16.2). Controls and cases were defined by a variability cutoff parameter. Cutoffs for the variability, which define the subset of positions with low or high variability, were chosen to split into balanced subsets with the most similar number of observations possible in each subset, *i.e.* using the median. Positions with higher variability than the cutoff are considered positives.

**Availability**

To ensure reproducibility and access to our results we provide at https://giancarlocroce.github.io/DCA_SARS-CoV-2/ the data generated in the course of this research and a jupyter-notebook to reproduce key figures and guide data analysis. This notebook will also contain data updated as compared to the datasets used in this article. The code to generate the predictions for the IND and DCA models is available at https://github.com/juan-rodriguez-rivas/covmut.

**Acknowledgments**

We are grateful to Erik Aurell, Matteo Bisardi and David Gfeller for interesting discussions, to David Gfeller in particular for his help with the immunologic data, and to Richard Neher for his valuable feedback on our manuscript. Our work was partially funded by the Faculty of Science and Engineering of Sorbonne University in the context of the call SU-COVID19-FSI, by the EU H2020 Research and Innovation Programme MSCA-RISE-2016 under Grant Agreement No. 734439 InferNet, and by the H2020 Marie Sklodowska Curie Individual Fellowship (H2020-MSCA-IF-2020), No.101027973 to Giancarlo Croce.

# Epistatic models predict mutable sites in SARS-CoV-2 proteins and epitopes

Juan Rodriguez-Rivas[1], Giancarlo Croce[2,3], Maureen Muscat[1], Martin Weigt*[1]

[1] Sorbonne Université, CNRS, Institut de Biologie Paris Seine, Computational and Quantitative Biology – LCQB, Paris, France.
[2] Department of Oncology, Ludwig Institute for Cancer Research Lausanne, University of Lausanne, Switzerland
[3] Swiss Institute of Bioinformatics (SIB), Lausanne, Switzerland

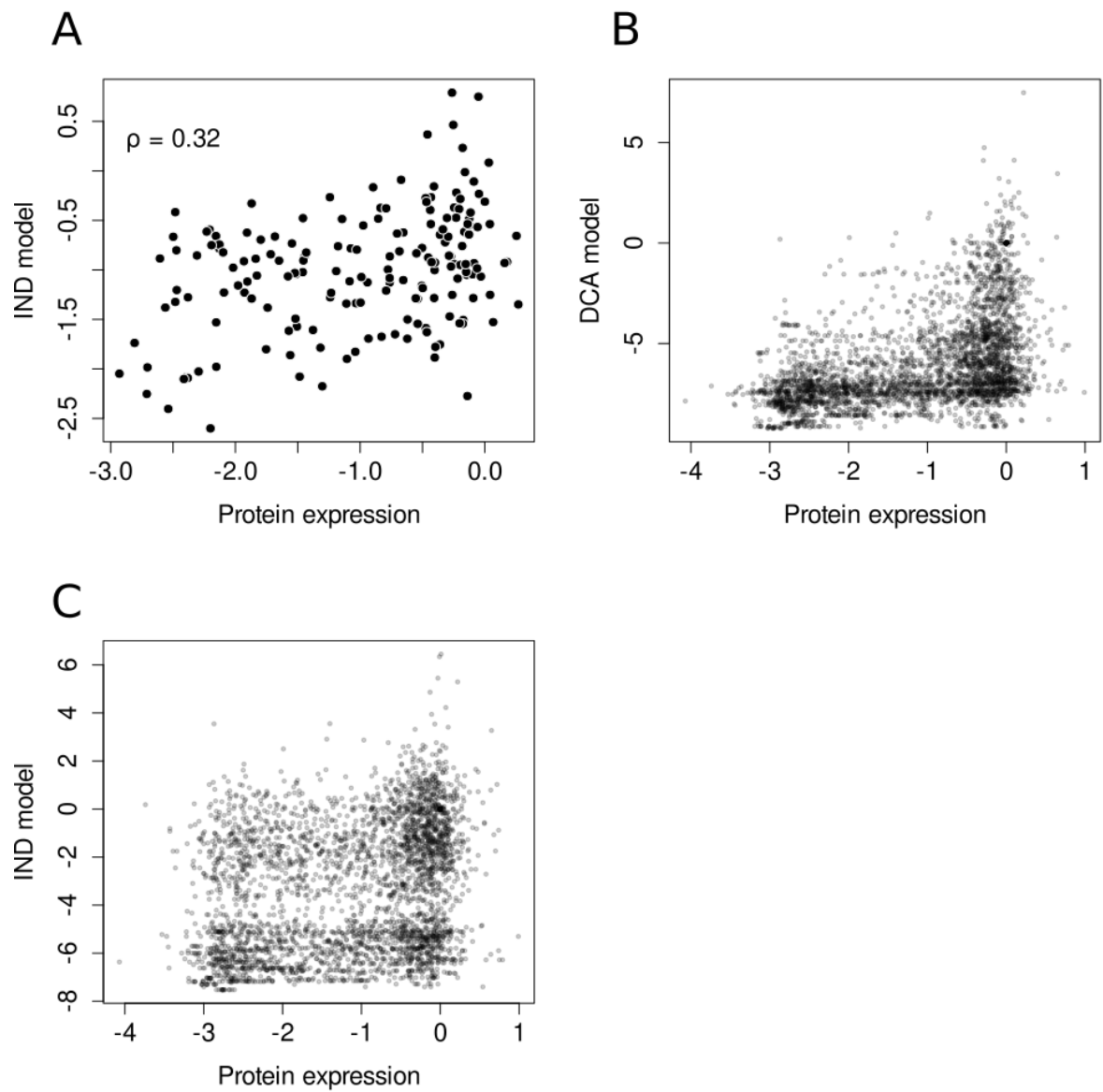

Fig. S1 A) Experimental protein expression for the 178 positions of the RBD as a function of the predicted effect by the IND model. The effects of 3355 single mutations in the 178 positions of the RBD measured by the experimental protein expression in the x-axis and predicted by the DCA model (B) and the IND model (C) in the y-axis.

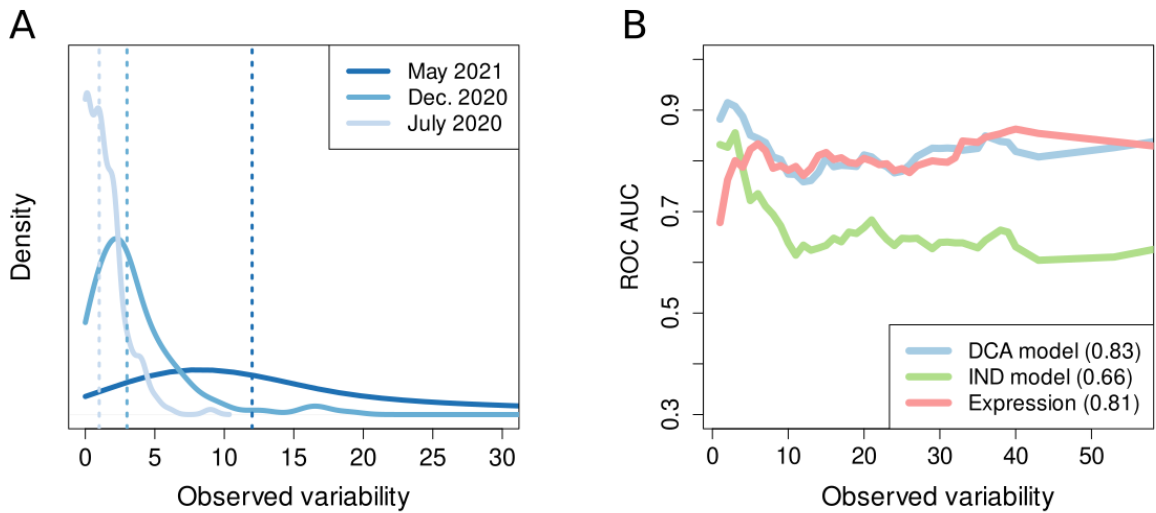

Fig. S2. A) Distributions of observed variability using the genomes available at July 2020, December 2020 and May 2021. The vertical dashed lines represent the median of each distribution and is used as a cutoff to distinguish between low- and high-variability positions. B) AUC for ROC curves using cutoffs of variability in the interval [1,56]. The mean AUC is shown in the legend.

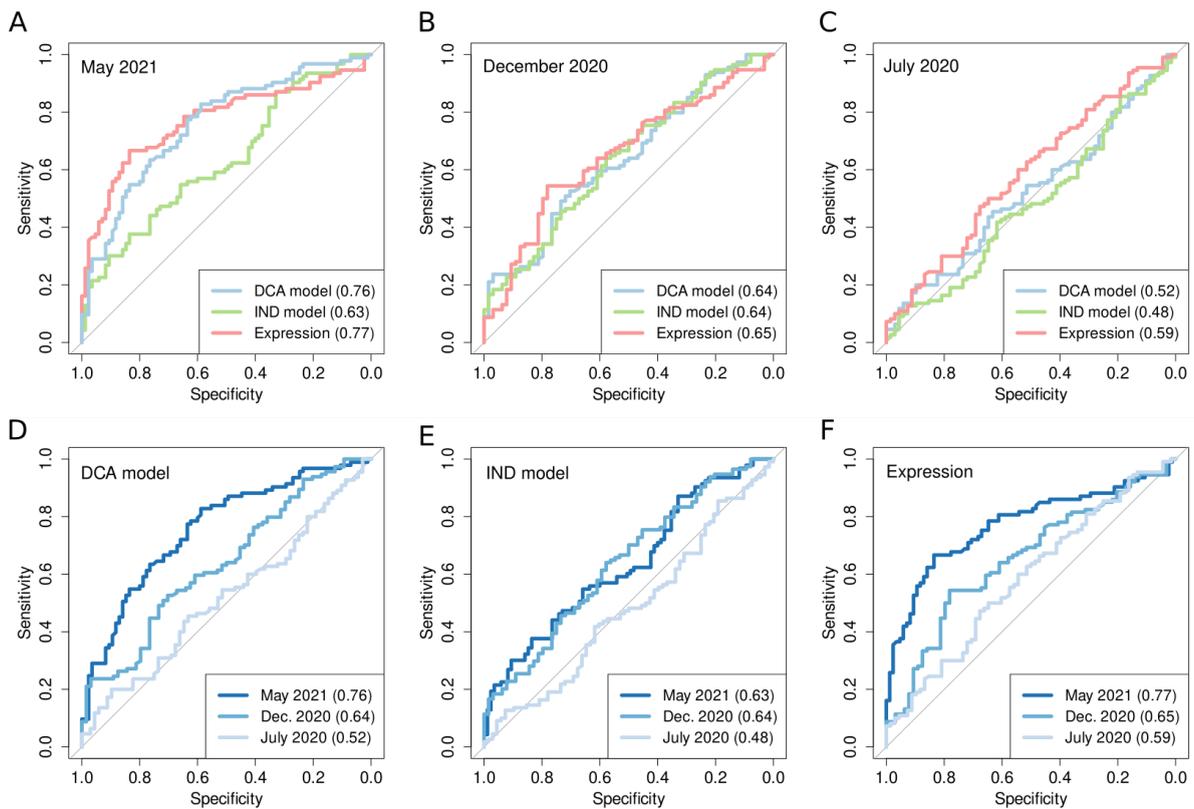

Fig S3. ROC curves for positions with low versus high observed variability for the 3 predictors with observed variability derived from all the SARS-CoV-2 genomes available at May 2021 (A), December 2020 (B), and July 2020 (C). ROC curves for positions with low versus high observed variability, where the observed variability is quantified with the SARS-CoV-2 genomes available at July 2020, December 2020, and May 2021 for the prediction coming from the DCA model (D), IND model (E), and protein expression (F).

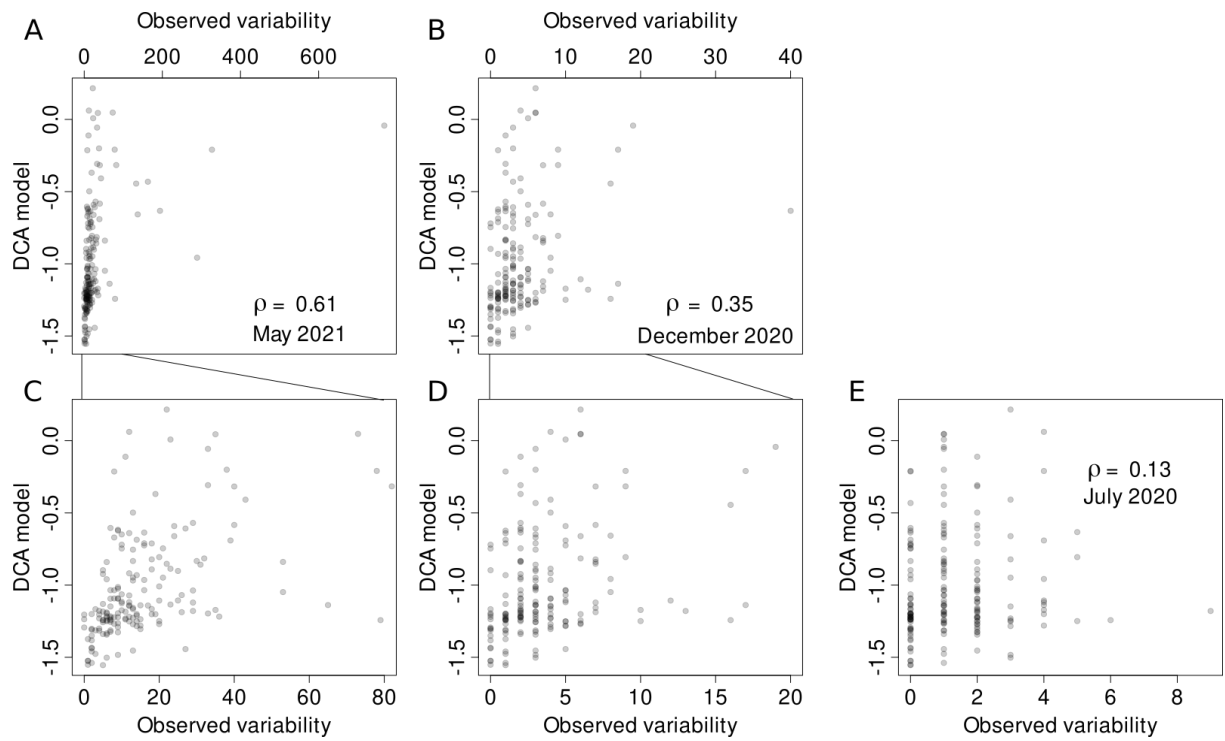

*Fig. S4. Observed variability (estimated with GISAID data till May 2021, panels A and C; December 2020, B and D; July 2020, E) compared to the DCA predictions. The whole range of observed variabilities is shown in panels A, B, and E. In panels C and D, a more restricted interval of observed variabilities is shown to improve the visibility.*

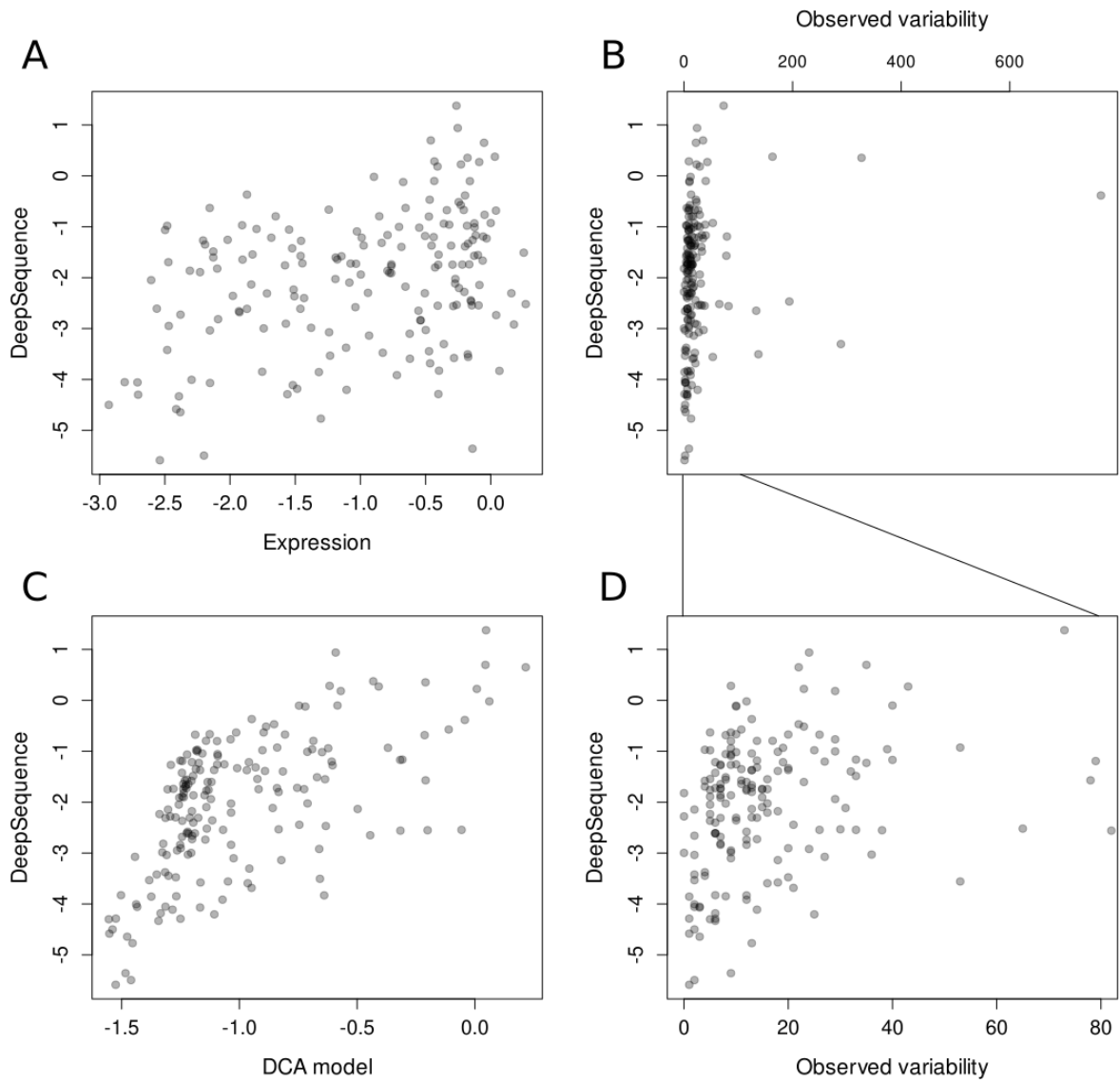

*Fig. S5. A) Scatter plot between protein expression and DeepSequence scores. Comparison between observed variability and DeepSequence scores for the whole range of observed variability in May 2021 (B) and in the interval of [0,80] (D). C) DeepSequence scores as a function of the DCA model scores.*

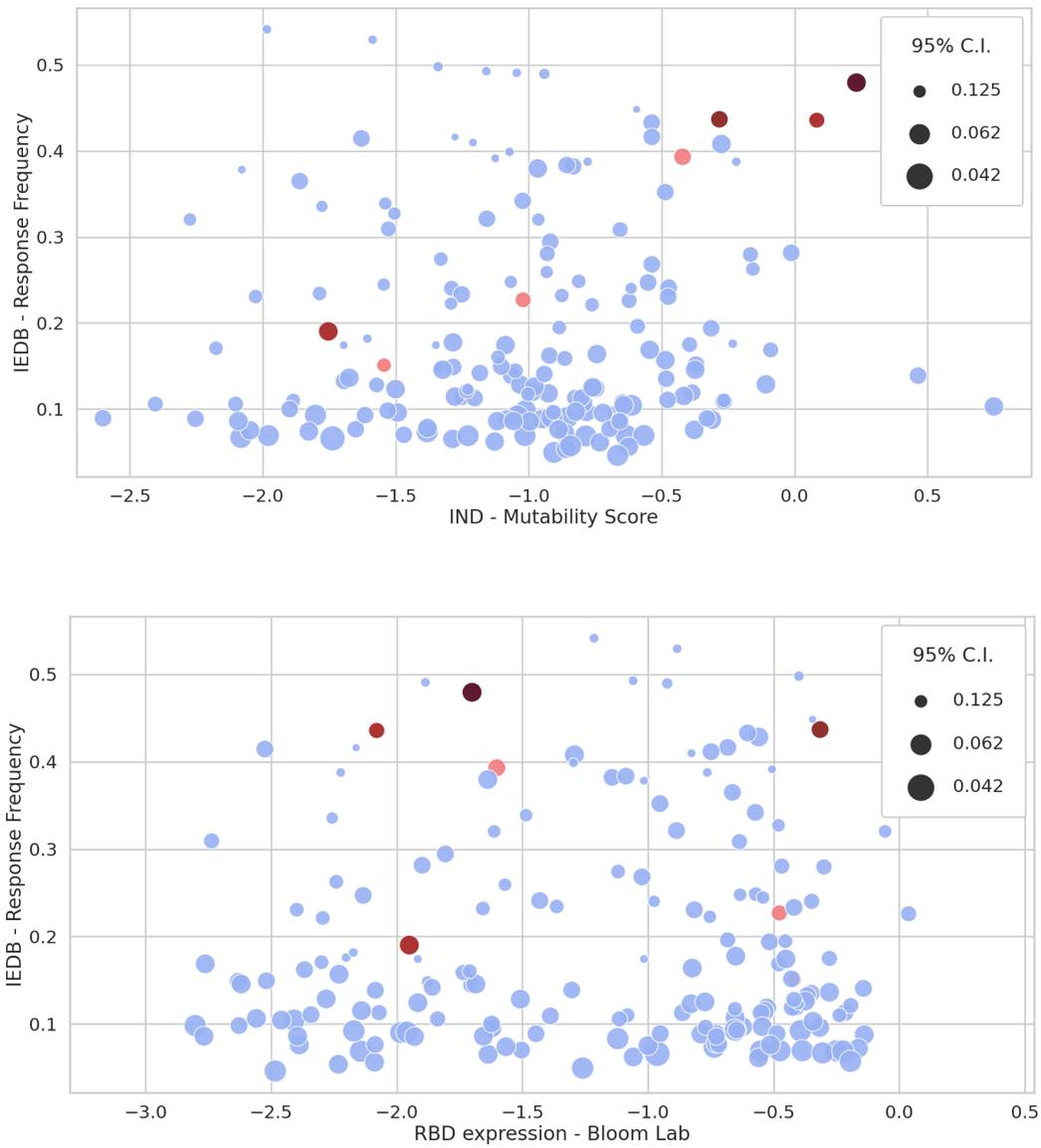

*Fig. S6. The IEDB-Response Frequency as a function of the IND mutability score (upper panel) or protein expression (lower panel) for each position of the RBD domain. The enrichment of VOC/VOI mutations becomes less pronounced as compared to the DCA score.*

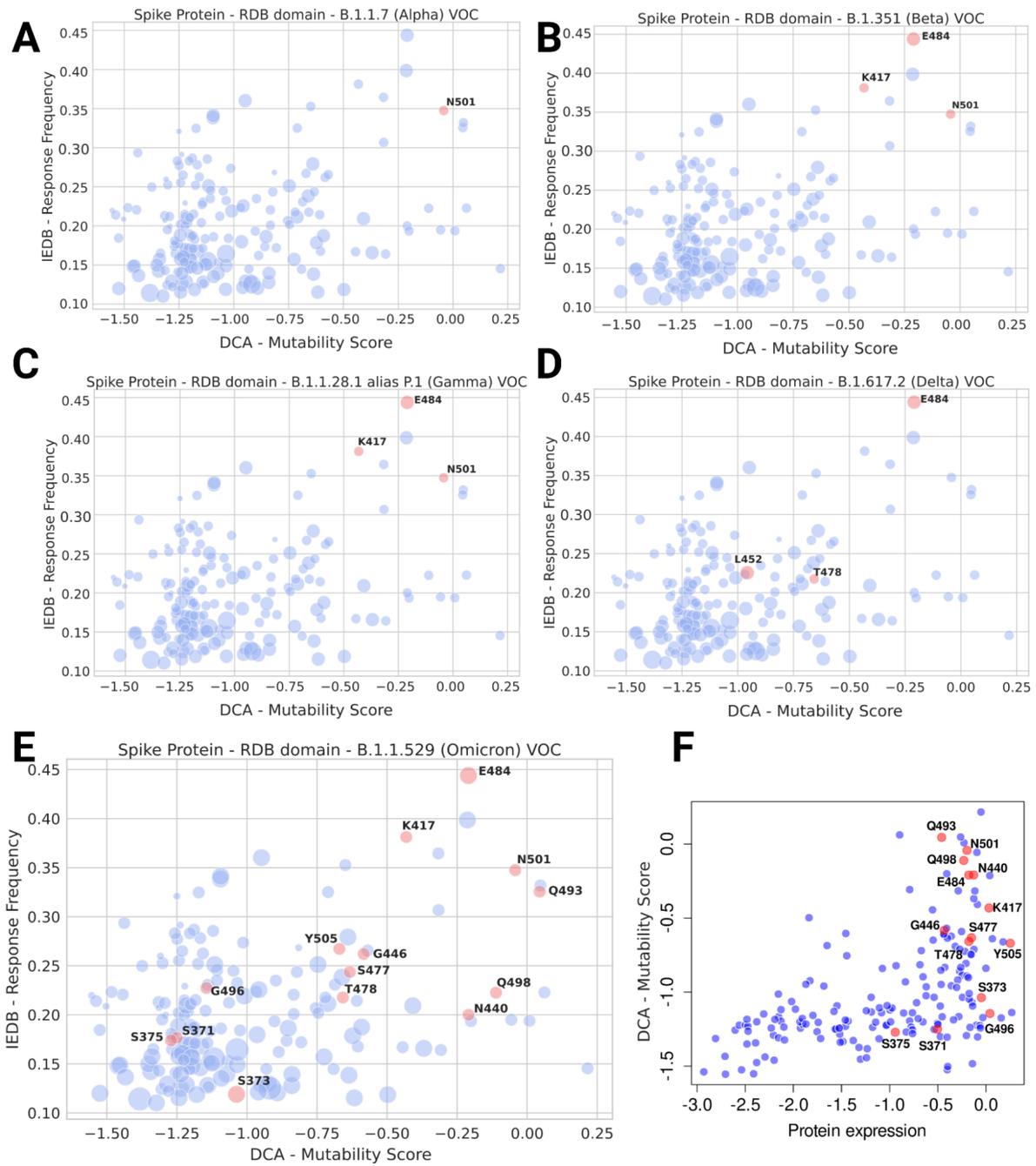

Fig. S7 - The IEDB-Response Frequency versus the DCA mutability score with updated IEDB response frequencies (data download 22 Nov 2021). We highlight in red the positions that are mutated in the 5 current VOCs, as of Dec 2021, indicated in https://cov-lineages.org/index.html: B.1.1.7 Alpha (Panel A), B.1.351 Beta (Panel B), P.1 Gamma (Panel C), B.1.617.2 Delta (Panel D), B.1.1.529 Omicron (Panel E) . We observe a pronounced enrichment for the Omicron variant in the upper right corner, i.e. positions that are likely to mutate (high DCA score), and whose mutations may cause immune escape (high IEDB - RF).  Interestingly,  the model predicts the positions S371, S373, S375, G496  - mutated in the Omicron variant (Panel E) -  to be deleterious, even if they are neutral in the expression experiments (Panel F). As discussed in the main text, this is likely to be due to (a) limited datasets of functional sequences or (b) mutations without effect on expression that may still be deleterious for overall protein fitness. Currently available data are not able to discriminate between these two possibilities.

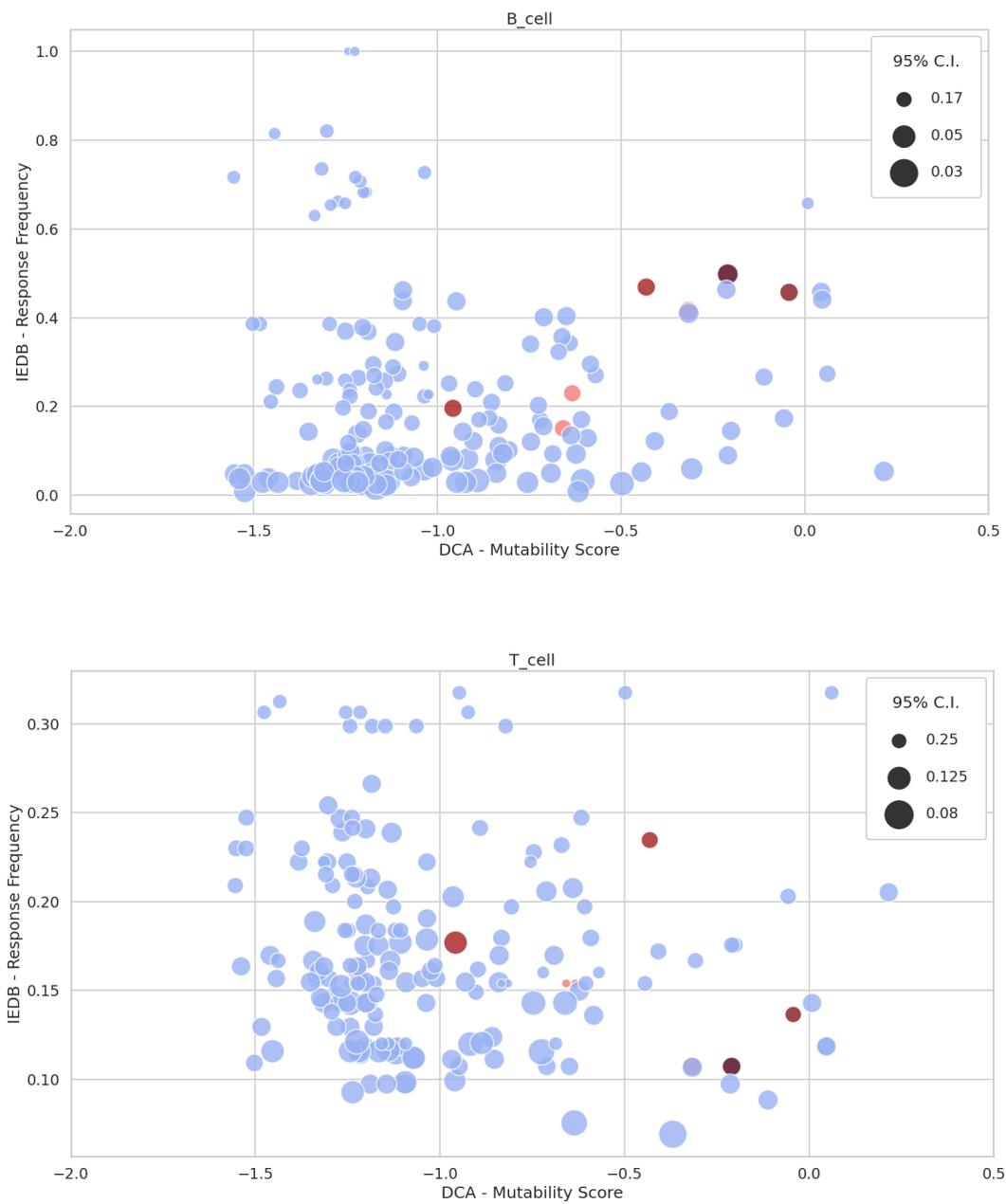

Fig S8. The IEDB-response frequency considering only B (upper panel) and T (lower panel) cell epitopes, and the DCA mutability score for each position of the RBD domain.

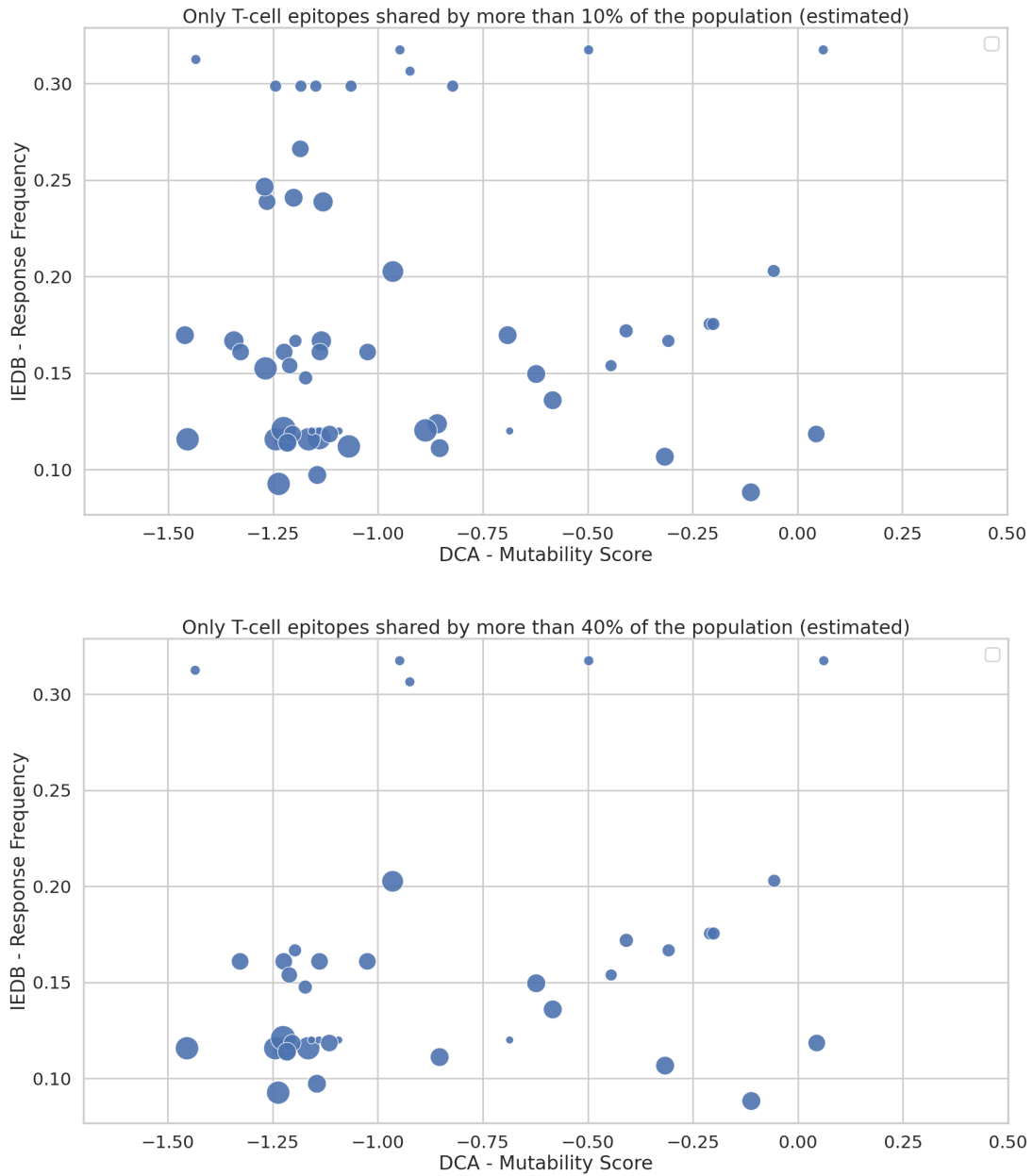

Fig S9. T cell immunoprevalent epitopes, i.e. predicted to be shared by at least the 10%(upper panel) and 40%(lower panel) of the world population. No clear correlation patterns between DCA and the IEDB emerge with the data available to date. Interestingly, no positions mutated in VOIs/VOCs were identified in T cell immunoprevalent epitopes.

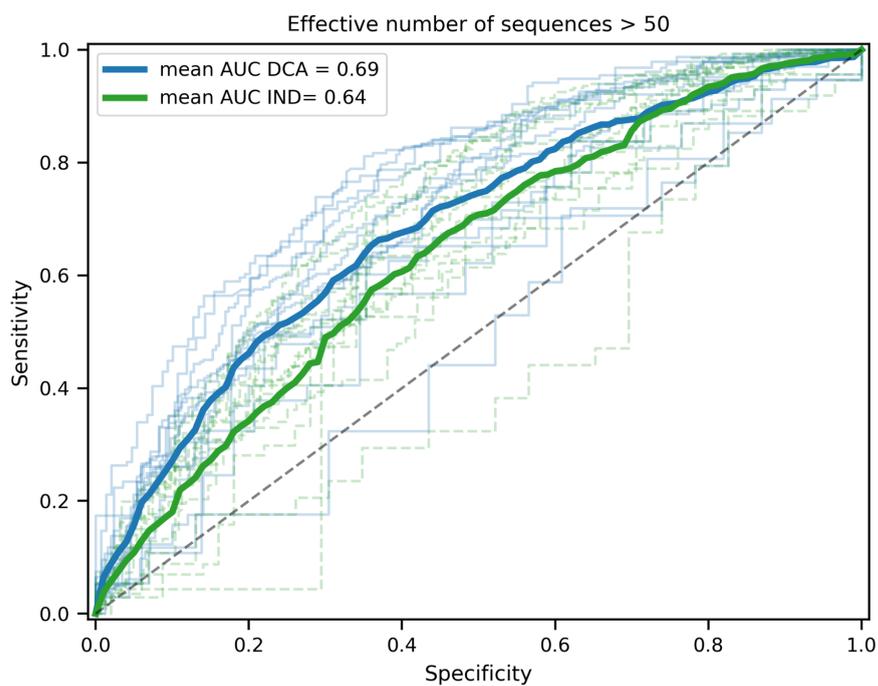

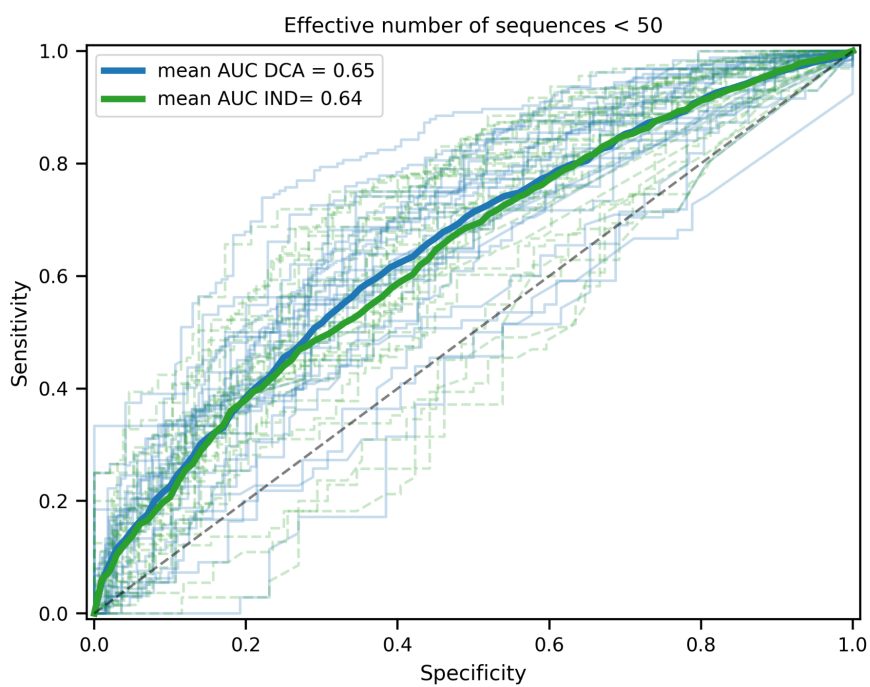

Fig S10. ROC curve for the DCA (blue) and IND (green) models for all 39 PFAM domains of the SARS-CoV-2 proteome, dividing between (upper panel) domains with more than 50 effective sequences (17 domains, 3491 positions) and (lower panel) less than 50 effective sequences (26 domains, 4546 positions). In bold, the mean ROC curves.

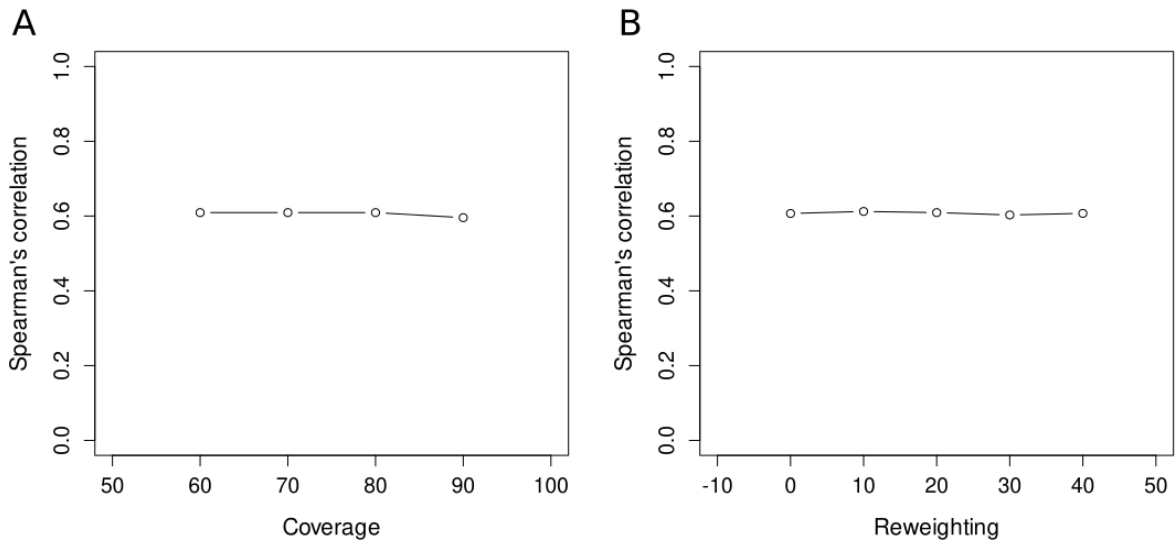

Fig S11. Spearman's correlation between the DCA model and the observed variability using different thresholds of coverage (filtering out sequences that do not cover that fraction of the reference sequence) (A) or the reweighting parameter (B) for model training.

*Table S1. List of Pfam protein domains in the SARS-CoV-2 proteome, and the number of effective sequences and positions in the corresponding MSA*

| Protein/ORF | Pfam identifier | Pfam accession | N. eff. seq. | N. positions |
|---|---|---|---|---|
| Envelope | CoV_E | PF02723.15 | 53 | 66 |
| Membrane | CoV_M | PF01635.19 | 40 | 201 |
| Nucleocapsid | CoV_nucleocap | PF00937.19 | 48 | 341 |
| ORF1a | bCoV_NAR | PF16251.6 | 19 | 98 |
| ORF1a | bCoV_NSP1 | PF11501.9 | 12 | 135 |
| ORF1a | bCoV_NSP3_N | PF12379.9 | 9 | 171 |
| ORF1a | bCoV_SUD_C | PF12124.9 | 2 | 64 |
| ORF1a | bCoV_SUD_M | PF11633.9 | 10 | 143 |
| ORF1a | CoV_NSP10 | PF09401.11 | 25 | 123 |
| ORF1a | CoV_NSP2_C | PF19212.1 | 22 | 167 |
| ORF1a | CoV_NSP2_N | PF19211.1 | 25 | 241 |
| ORF1a | CoV_NSP3_C | PF19218.1 | 58 | 488 |
| ORF1a | CoV_NSP4_C | PF16348.6 | 33 | 96 |
| ORF1a | CoV_NSP4_N | PF19217.1 | 58 | 354 |
| ORF1a | CoV_NSP6 | PF19213.1 | 71 | 262 |
| ORF1a | CoV_NSP7 | PF08716.11 | 31 | 83 |
| ORF1a | CoV_NSP8 | PF08717.11 | 30 | 197 |
| ORF1a | CoV_NSP9 | PF08710.11 | 33 | 113 |
| ORF1a | CoV_peptidase | PF08715.11 | 103 | 319 |
| ORF1a | Macro | PF01661.22 | 10075 | 107 |
| ORF1a | Peptidase_C30 | PF05409.14 | 41 | 291 |
| ORF1b | CoV_Methyltr_1 | PF06471.13 | 27 | 522 |
| ORF1b | CoV_Methyltr_2 | PF06460.13 | 31 | 296 |
| ORF1b | CoV_NSP15_C | PF19215.1 | 42 | 153 |
| ORF1b | CoV_NSP15_M | PF19216.1 | 40 | 97 |
| ORF1b | CoV_NSP15_N | PF19219.1 | 38 | 61 |
| ORF1b | CoV_RPol_N | PF06478.14 | 31 | 352 |
| ORF1b | RdRP_1 | PF00680.21 | 61 | 489 |
| ORF1b | Viral_helicase1 | PF01443.19 | 81660 | 225 |
| ORF3a | bCoV_viroporin | PF11289.9 | 3 | 274 |
| ORF6 | bCoV_NS6 | PF12133.9 | 4 | 61 |
| ORF7a | bCoV_NS7A | PF08779.11 | 8 | 106 |
| ORF7b | bCoV_NS7B | PF11395.9 | 3 | 42 |
| ORF8 | bCoV_NS8 | PF12093.9 | 4 | 118 |
| Spike | bCoV_S1_N | PF16451.6 | 103 | 305 |
| Spike | bCoV_S1_RBD | PF09408.11 | 83 | 178 |
| Spike | CoV_S1_C | PF19209.1 | 6231 | 57 |
| Spike | CoV_S2_C | PF19214.1 | 50 | 40 |
| Spike | CoV_S2 | PF01601.17 | 79 | 522 |

Table S2. Pfam domains and the number of effective sequences in the MSAs obtained starting with full-length protein sequence or the domain sequence. In bold, the domains where there is a substantial difference.

| Protein/ORF | Pfam identifier | N. eff. seqs. full-length | N. eff. seqs. domain |
|---|---|---|---|
| Envelope | CoV_E | 53 | 49 |
| Membrane | CoV_M | 40 | 37 |
| Nucleocapsid | CoV_nucleocap | 48 | 47 |
| ORF3a | bCoV_viroporin | 3 | 3 |
| ORF6 | bCoV_NS6 | 4 | 4 |
| ORF7a | bCoV_NS7A | 8 | 8 |
| ORF7b | bCoV_NS7B | 3 | 3 |
| ORF8 | bCoV_NS8 | 4 | 4 |
| **Spike** | **bCoV_S1_N** | **103** | **48** |
| **Spike** | **bCoV_S1_RBD** | **83** | **26** |
| **Spike** | **CoV_S1_C** | **123** | **6231** |
| **Spike** | **CoV_S2_C** | **50** | **7** |
| Spike | CoV_S2 | 78 | 79 |

Table S3. Strongest inter-domain epistatic couplings for pairs of domains with a maximum (out of all the possible inter-domain epistatic couplings between the pair of domains) coupling higher than 0.5.

| First protein/ORF | First domain | Second protein/ORF | Second domain | Max. coupling |
|---|---|---|---|---|
| ORF1a | CoV_NSP2_N | ORF3a | bCoV_viroporin | 0.85 |
| ORF3a | bCoV_viroporin | ORF8 | bCoV_NS8 | 0.63 |
| ORF1a | CoV_RPol_N | Spike | bCoV_viroporin | 0.57 |
| ORF1b | CoV_NSP2_N | ORF3a | bCoV_S1_RBD | 0.56 |

# SI text

**Sequence data**

Sequence data in FASTA format were downloaded from the following databases: GISAID (release 16 May 2021), Uniref90 (ref, release December 2020), ViPR (downloaded in September 2020), NCBI viral genomes (downloaded in September 2020) and MERS coronavirus database ( downloaded in September 2020). The amino acid sequence of isolate Wuhan-Hu-1 was used as the reference proteome (genbank identifier: MN908947). Protein domains were detected using the HMMER suite (ref, version 3.1b2) and the HMM profiles from Pfam. After running the command *hmmsearch* from the HMMER suite (ref, version 3.1b2) on the reference proteome using the HMM profiles of SARS-CoV-2 provided by Pfam, the domain amino acid sequence of the full-length protein were trimmed accordingly to the *hmmsearch* output to obtain a reference sequence for each domain. We kept all non-overlapping pfam domains with a domain e-value lower than $10^{-5}$.

A global sequence database including distant species was built by combining Uniref90, ViPR, NCBI viral genomes and MERS coronavirus database. Starting with the domain sequences, we built MSAs by running *jackhmmer* with 5 iterations. For the proteins not belonging to the ORF1ab (which is too long to apply this procedure), we also built MSAs with *jackhmmer* with 5 iterations starting with the full-length reference protein sequences instead of domain sequences. The resulting full-length protein MSAs were decomposed and trimmed to domain alignments by keeping the corresponding columns. When two MSAs from the global database were obtained (one coming from the full-length sequence and another coming from the domain sequence), the one with the highest number of sequences non-redundant at 80% was kept for further analysis for each Pfam domain. Although both strategies usually recover a similar number of sequences, there exists a substantial difference for most domains in the Spike protein (Table S2), allowing us to increase the available sequence data in their MSAs. As quality controls, all sequences including non-standard amino acids were removed as well as repeated sequences or sequences covering less than 80% of the reference domain sequence. To avoid a bias toward the reference sequence, all sequences closer than 90% sequence identity to the Wuhan-Hu-1 reference were filtered out.

For the GISAID database, an MSA for each domain sequence was built using the command *jackhmmer* from the HMMER suite with only 1 iteration as the GISAID sequences are very similar to those in the reference proteome. We filtered sequences including non-standard amino acids, coverage lower than 80%, and those belonging to a non-human host. Only non-identical sequences were considered to avoid the strong sequencing bias due to the highly diverse number of genomes sequenced in different countries. The variability of each position was estimated by counting the number of sequences that have a different amino acid in the corresponding position compared to the reference.

**Co-alignments of domains**

Starting with the 39 raw alignments of domains constructed using both the alignments of distant and close sequences (*Materials and methods*), we build 741 co-alignments (all possible combinations of two domains) by joining the sequences coming from the same genome (thanks to the genome accession number). Co-alignments with fewer than 50 effective sequences were discarded to increase the reliability of the predictions. Note that the number of effective sequences is higher in this analysis compared to the mutability predictions because of the large number of close sequences. From the remaining 601 co-alignments, we computed the models as in case of single domains (see *Material and methods*) and obtained the APC scores from the DCA model between each pair of inter-domains positions. The pairs of domains with at least one APC score higher than 0.3 are linked in Fig. 4D. The 4 predictions with the highest APC scores can be found in Table S3.